\documentclass[
 reprint,
superscriptaddress,
 amsmath,amssymb,
 aps,
 prr
]{revtex4-2}
\usepackage[T1]{fontenc}
\usepackage{amsmath}
\usepackage[]{algorithm2e}
\usepackage{tikz}
\usetikzlibrary{shapes,arrows,positioning}
\usepackage{pgfplots}
\tikzset{decision/.style={diamond, draw, fill=yellow!20, 
    text width=10em, aspect=2, text badly centered, node distance=3cm, inner sep=0pt, minimum height=4em},
block/.style={rectangle, draw, fill=gray!20, 
    text width=15em, text badly centered, rounded corners, minimum height=4em},
small_block/.style={circle, draw, fill=white!20, 
    text width=0em, text centered, rounded corners, minimum height=0em},    
line/.style={draw, -latex'},
cloud/.style={draw, ellipse,fill=red!20, node distance=3cm,text width=15em,
    minimum height=6em, text centered,},}

\usepackage[caption=false]{subfig}

\usepackage{comment}
\usepackage{graphicx}
\usepackage{dcolumn}
\usepackage{bm}
\usepackage{xcolor}
\usepackage{soul}
\usepackage{ulem}
\usepackage[hidelinks,colorlinks=true,linkcolor=blue,citecolor=blue]{hyperref}

\include{macros}

\usepackage{cancel}

\begin{document}

\preprint{}

\title{\articleTitle}

\author{Jacob Knight}
\email{jwk21@ic.ac.uk}
\affiliation{Department of Mathematics, Imperial College London, South Kensington, London SW7 2BZ, UK}

\author{Farid Kaveh}
\affiliation{Department of Mathematics, Imperial College London, South Kensington, London SW7 2BZ, UK}

\author{Gunnar Pruessner}
\email{g.pruessner@imperial.ac.uk}
\affiliation{Department of Mathematics, Imperial College London, South Kensington, London SW7 2BZ, UK}

\date{20 July 2025}

\begin{abstract}
Entropy production distinguishes equilibrium from non-equilibrium. 
Calculating the entropy production rate (EPR) is challenging in systems where some degrees of freedom cannot be observed.
Here we introduce a perturbative framework to calculate the ``partial EPR''  of a canonical hidden-state system, a generic self-propelled active particle with hidden self-propulsion. We find that the parity symmetry, $\SymParity$, and (time-)reversibility, $\SymTime$, of the hidden variable determine partial entropy production. Non-trivial entropy production appears at least at \emph{sixth order} in the self-propulsion velocity. We apply our framework to two processes which break $\SymParity$- and $\SymTime$-symmetries respectively: an asymmetric telegraph process and diffusion with stochastic resetting.
\end{abstract}

\maketitle

\begin{figure}[t]
\centering
\includegraphics[width=1.\linewidth]{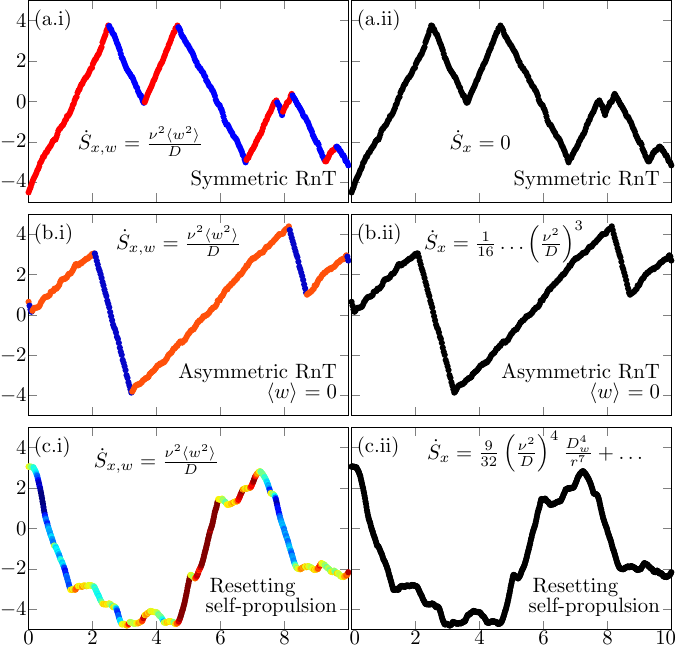}
\caption{Illustration of three processes with (i) fully resolved and (ii) hidden self-propulsion velocity, in the left and right columns respectively. Rows correspond to self-propulsion velocities undergoing: a) a symmetric telegraph process (symmetric RnT), b) an asymmetric telegraph process with vanishing net drift $\Wave{w}=0$ (asymmetric RnT), \Eref{asym_EPR_without_drift}, c) diffusion with stochastic resetting (Resetting self-propulsion), \Eref{resetting_epr}. In the left-hand column, red and blue colouring correspond to positive and negative self-propulsion velocity $w(t)$ respectively. In a) trajectories break time-reversal symmetry (TRS) when fully observed (since blue and red colours are reversed) but obey TRS when the self-propulsion state is hidden. In b) and c) trajectories break TRS even with hidden $w(t)$. 
All processes are discussed below, also \Tref{processes}.
    }
    \flabel{process_illustrations}
\end{figure}

Violation of time-reversal symmetry (TRS) is a fundamental feature of non-equilibrium systems and essential for the existence of life \cite{montrollStatisticalMechanicsTransport1954,astumianBrownianMotors2002,michaelianEntropyProductionOrigin2011}. The extent of TRS-breaking places thermodynamic constraints on biological processes such as reproduction, sensing and signalling \cite{thurleyReliableEncodingStimulus2014,lanEnergySpeedAccuracy2012,harveyUniversalEnergyaccuracyTradeoffs2023,englandStatisticalPhysicsSelfreplication2013}. A pressing theoretical challenge is to estimate or bound the time-irreversibility of systems which \emph{cannot be fully observed}, since visible degrees of freedom may break TRS to a lesser extent than the complete underlying system \cite{dieballPerspectiveTimeIrreversibility2025}. When \emph{all} microscopic details are visible, time-irreversibility is measured by the entropy production rate (EPR), which may intuitively be thought of as a ``distance" of the system from equilibrium \cite{FodorETAL:2016}. Zero EPR corresponds to a time-reversible (i.e. equilibrium) system. A number of approaches have been proposed for partially hidden systems: for discrete state systems, lower bounds to the full entropy production rate have been derived from waiting time distributions and transition frequencies \cite{biskerHierarchicalBoundsEntropy2017, nitzanUniversalBoundsEntropy2023, martinezInferringBrokenDetailed2019, vandermeerThermodynamicInferencePartially2022, ertelOperationallyAccessibleUncertainty2022, deguntherFluctuatingEntropyProduction2024} and estimates proposed based on inference of hidden system parameters \cite{degunthervandermeerseifertcalmodulin}. In continuous systems, approaches to quantifying irreversibility include ``milestoning" continuous trajectories into discrete states \cite{blomMilestoningEstimatorsDissipation2024, hartichViolationLocalDetailed2023} and explicit integration for particles with hidden Gaussian (Ornstein-Uhlenbeck) activity \cite{FodorETAL:2016,dabelowIrreversibilityActiveMatter2019,capriniEntropyProductionOrnstein2019}.

In this Letter we calculate a ``partial" EPR to quantify the TRS-breaking of a canonical partially observable system: an overdamped active particle with hidden stochastic self-propulsion \cite{power_extraction, DabelowBoEichhorn:2021, dabelowIrreversibilityActiveMatter2019, capriniEntropyProductionOrnstein2019}. Our perturbative framework allows us to characterise \emph{generic} active particles, going beyond the previously studied case of active Ornstein-Uhlenbeck particles \cite{FodorETAL:2016,dabelowIrreversibilityActiveMatter2019,capriniEntropyProductionOrnstein2019}. Intriguingly, we find that the \textit{symmetries} of the hidden degree of freedom determine whether or not, and the extent to which, the TRS of spatial trajectories is broken. We demonstrate this by calculating the leading-order partial EPR of two processes which minimally break the required symmetries to observe TRS-breaking in their spatial trajectories.

Intuition is provided by considering a symmetric run-and-tumble (RnT) particle \cite{TailleurCates:2008,CatesTailleur:2013} moving in one dimension, \Fref[a]{process_illustrations}. When both its spatial position and self-propulsion velocity can be observed, trajectories can easily be distinguished from their time-reversed counterparts, meaning that the process has a non-zero EPR. However, when \textit{only} its position can be observed, trajectories are time-reversal symmetric and thus the process has zero \textit{partial} EPR. \Frefs[b,c]{process_illustrations} show processes where entropy production is reduced but non-zero when the self-propulsion velocity is hidden, \textit{i.e.} processes with non-zero partial EPR.

\textit{Key derivation: }
In the following, the Langevin equation \cite{TailleurCates:2008,CatesTailleur:2013} 
\begin{subequations}
\elabel{langevin}
\begin{align}
        \xdot(t)&= \nu w(t)+\xi(t)\;,\\
        \XIave{\xi(t) \xi(t')} &= 2 D \delta(t-t')\;, \elabel{noise_correlator} 
\end{align}
\end{subequations}
acts as a phenomenological description of an overdamped active particle at position $x(t)\in[0,L)$ with periodic boundary conditions, as to enforce a steady state. It is
self-propelled with velocity $\nu w(t)$, where $\nu$ has dimensions of velocity and $w(t)$ is a steady-state dimensionless stochastic process, to be specified further, but evolving independently of $x(t)$. The particle is further subject to independent Gaussian white noise $\xi(t)$ with diffusivity $D$ and vanishing mean, $\XIave{\xi}=0$, where $\XIave{\bullet}$ denotes the average over the noise. 
The total steady-state internal EPR of \Eref{langevin} depends on the ratio of the path probabilities of forward and reversed trajectories of $\{x(t)\}$ and $\{w(t)\}$ and is given by \cite{gaspard2004time,cocconi2020entropy,Sekimoto:2010}
\begin{multline}\elabel{EP_full}
    \EPRfull = \lim_{T \to \infty} \frac{1}{T} \int\Dint{x}\!\!\int\Dint{w}\,\PprobXW[\xpath,\wpath] \\ 
    \times\ln\left(\frac{\PprobXW[\xpath, \wpath]}{\PprobXW[\Rxpath, \Rwpath]}\right)\;,
\end{multline}
where $T$ is the trajectory duration. Trajectories are assumed to be initialised according to steady-state distributions, negating transient contributions arising from initial conditions (which vanish anyway due to the limit $T \to \infty$). 
$\PprobXW[\xpath,\wpath]$ denotes a normalised measure for the steady-state probability density of a joint path $[\xpath,\wpath]$ for $t\in[0,T]$ to occur in process \Eref{langevin}. Correspondingly, $\PprobXW[\Rxpath,\Rwpath]$ is the probability density for the reverse path to occur. Such path probabilities are easy to construct when they derive from a known distribution, \eg using an Onsager-Machlup functional based on \Eref{langevin} for the path probability \cite[][Ch.~5.2.2.3]{Taeuber:2014, roldanMartingalesPhysicistsTreatise2023}
\begin{equation}\elabel{OM}
    \PprobXW[\xpath|\wpath]\propto \Exp{-\frac{1}{4D}\int_0^T \!\!\!\dint{t}\,(\xdot(t)-\nu w(t))^2}\;,
\end{equation}
of $\xpath$ \emph{conditioned} on $\wpath$, derived from the density of the noise $\xipath$. To avoid ambiguity, we use the Stratonovich convention throughout. Here and in the following, $\xdot(s)$ shall denote the derivative of $x(t)$ with respect to $t$, evaluated at $s$, so that
\begin{multline}\elabel{OM_reversed}
    \PprobXW[\Rxpath|\Rwpath]\\
    \propto \Exp{-\frac{1}{4D}\int_0^T \!\!\!\dint{t}\,(-\xdot(T-t)-\nu w(T-t))^2}\;.
\end{multline}
Multiplying \Eref{OM} by $\PprobW[\wpath]$, which generally is difficult to obtain, finally produces $\PprobXW[\xpath,\wpath]$. Instead of using such an explicit expression in \Eref{EP_full}, one may use that $x(t),w(t)$ is Markov, so that $\PprobXW[\xpath,\wpath]$ factorises in time-slices \cite{gaspard2004time}
and the limit of large $T$ does not need to be taken, reducing the ratio of path probabilities to that of transition rates \cite{cocconi2020entropy,PruessnerGarcia-Millan:2022}.

Such a bypass is no longer available if
we consider the scenario where the position $x(t)$ of the particle can be observed, but \emph{not} its internal self-propulsion state $w(t)$. The postion $x(t)$ without accompanying $w(t)$ leaves the system non-Markovian. This \emph{partial} description of the system has a corresponding \emph{partial} EPR
\begin{equation}\elabel{partial_EP_defn}
    \EPRpartX = \lim_{T \to \infty} \frac{1}{T} \int \Dint{x}\;\PprobX[\xpath]\ln\left(\frac{\PprobX[\xpath]}{\PprobX[\Rxpath]}\right)\;,
\end{equation}
which by Jensen's inequality can be shown to be bounded from above by $\EPRfull\ge\EPRpartX$. 
Our approach to calculate \Eref{partial_EP_defn} is to express the path probability as 
\begin{equation}\elabel{Pprobx_from_conditional}
    \PprobX[\xpath] =\int \Dint{w} \PprobXW[\xpath|\wpath]\PprobW[\wpath] \;,
\end{equation}
where the dummy variable$w(t)$ can equally be replaced by $w(T-t)$.
\begin{widetext}
Using \Erefs{OM}, \eref{OM_reversed} and \eref{Pprobx_from_conditional} in \Eref{partial_EP_defn} allows the term quadratic in $\dot{x}(t)$ to be cancelled across the fraction as $x(t)$ is a dummy variable of the outer path integral. Using that $\plaind x(t)/\plaind t=-\xdot(T-t)$,
the partial EPR takes the form 
\begin{gather}\elabel{partial_EP_3}
    \EPRpartX = \lim_{T \to \infty} \frac{1}{T} \int \Dint{x} \PprobX[\xpath]
    \ln\left(\frac{\WSTARave{\Exp{\frac{\nu}{2D} \int_0^T \dint{t} w(t) \dot{x}(t)}}}{\WSTARave{\Exp {-\frac{\nu}{2D} \int_0^T \dint{t} w(T-t) \dot{x}(t)}}}\right)\\
\text{with }\qquad 
\elabel{overline_def_main}
    \WSTARave{\bullet}
    =\frac{1}{\NormWSTAR(\nu)}\int\Dint{w}\bullet 
    \Exp{-\frac{\nu^2}{4D} \int_0^T\dint{t} w(t)^2 }
    \;\PprobW[\wpath] =
    \int\Dint{w}\bullet \PprobWSTAR[\wpath] \;,
\end{gather}
where $\NormWSTAR(\nu)$ is the normalisation of the modified path probability distribution $\PprobWSTAR[\wpath]$ defined above. 
To leading order $\nu^0$ of the exponential, $\WSTARave{\bullet}$ is the expectation $\Wave{\bullet}$ under $\PprobW[\wpath]$, \SMref{SM:notation}.
The logarithm of the expectation of an exponential in \Eref{partial_EP_3} can be recognised as the cumulant-generating function of $\PprobWSTAR$ with $\dot{x}(t)$ as the conjugate variable. Writing these expectations as
\end{widetext}
\begin{equation}\elabel{xdot_barred}
    \Xave{\xdot(t_1)\sdots\xdot(t_n)} = 
    \int\Dint{x} \Big(\xdot(t_1)\sdots\xdot(t_n)\Big)\  \PprobX[\xpath]
\end{equation}
\Eref{partial_EP_3} can be rewritten in terms of cumulants and correlators,
\begin{multline}\elabel{partial_EP_expansion_free}
    \EPRpartX 
    = \lim_{T \to \infty} \frac{1}{T} \sum_{n=1}^\infty \frac{1}{n!}\left(\frac{\nu}{2D}\right)^n \int_0^T \dint{t_1}\!\!\sdots\dint{t_n}
    \ \Xave{\xdot(t_1)\sdots\xdot(t_n)}
    \\ 
    \times \Big(\WSTARaveS{w(t_1)\sdots w(t_n)}{\cbb}
    -(-)^{n}\WSTARaveS{w(T-t_1)\sdots w(T-t_n)}{\cbb}\Big) \;,
\end{multline}
where the superscripted $\cbb$ indicates the cumulants. \Eref{partial_EP_expansion_free} is central to the results derived in the following. By suitable substitution, $t_n'=T-t_n$, the integrand can be symmetrised with respect to moments and cumulants, \Eref{partial_EP_expansion_symmetric}.

\textit{Results: }
Analysing \Eref{partial_EP_expansion_free}, we observe that the partial EPR of processes of the form \Eref{langevin} depends on the \emph{symmetries} obeyed by the hidden self-propulsion $w(t)$. The relevant symmetries are parity symmetry $\SymParity$, which is obeyed by $\PprobW[\wpath]$ iff
\begin{equation}
    \PprobW[\wpath]=\PprobW[\Mwpath]=\PprobW[\SymParity\wpath]
\end{equation}
and time-reversal symmetry $\SymTime$
\begin{equation}
    \PprobW[\wpath]=\PprobW[\Rwpath]=\PprobW[\SymTime\wpath] \ .
\end{equation}
By construction of \Erefs{EP_full} and \eref{partial_EP_defn}, entropy production vanishes
if $\PprobXW[\xpath,\wpath]$ or $\PprobX[\xpath]$ respectively
are $\SymTime$-symmetric. As $(x-1)\ln(x)\ge0$ for $x\in \Rset^+$ the converse holds as well.

The $\SymParityTime$-symmetry of $\int_0^T\dint{t}w^2(t)$ in \Eref{overline_def_main} implies that the relevant symmetries of $\PprobWSTAR[\wpath]$ are identical to those of $\PprobW[\wpath]$, so we focus on the latter.
By inspection of \Eref{partial_EP_3}, if $\PprobW[\wpath]$ is invariant under replacing $\wpath$ by $\RMwpath$, then the logarithm vanishes, so that $\EPRpartX=0$ under $\SymParityTime$-symmetry of $w(t)$ \cite{power_extraction}. Given $(x-1)\ln(x)\ge0$, the converse follows from \Eref{partial_EP_3}, \ie a process \Eref{langevin} with vanishing partial EPR must be driven by $\SymParityTime$-symmetric $w(t)$.

Further, the expansion \Eref{partial_EP_expansion_free} reveals separate cancellation of  odd and even terms under $\SymParity$ and $\SymTime$ symmetry, respectively. If $\PprobW[\wpath]$ is $\SymParity$-symmetric, then any odd moments and cumulants of $w(t)$ vanish, \ie odd $n$ vanish in \Eref{partial_EP_expansion_free}. If $\PprobW[\wpath]$ is $\SymTime$-symmetric, 
$\WSTARave{w(t_1)\sdots w(t_n)}=\WSTARave{w(T-t_1)\sdots w(T-t_n)}$, then even $n$ cancel.

The lowest order $n=1$ in \Eref{partial_EP_expansion_free} indicates ``trivial'' entropy production as $x(t)$ displays a net drift, $\WSTARave{w}=\WSTARaveS{w}{\cbb}\ne 0$, as we will illustrate below. The second lowest order, $n=2$, always vanishes in the steady state, as $\WSTARaveS{w(t_1)w(t_2)}{\cbb}$ can be a function only of $|t_1-t_2|$ and therefore equals $\WSTARaveS{w(T-t_1)w(T-t_2)}{\cbb}$.
This immediately explains that free active Ornstein-Uhlenbeck particles have vanishing partial entropy production, $\EPRpartX=0$, as all cumulants beyond $n=2$ of $\PprobW[\wpath]$ and $\PprobWSTAR[\wpath]$ vanish. More terms need to be considered in the presence of an external potential \cite{KnightPruessner:2025b}, resulting in vanishing partial entropy production in harmonic, but generally non-zero partial EPR in anharmonic potentials  \cite{DabelowBoEichhorn:2021}.

After symmetrising \Eref{partial_EP_expansion_free}, the leading order contribution to $\EPRpartX$ can be written as (\SMref{calc_EPR})
\begin{multline}\elabel{EPR_leading_order_main}
    \EPRpartX=\lim_{T \to \infty} \frac{1}{2T} \frac{1}{\nlowest !}\left(\frac{\nu^2}{2D}\right)^{\nlowest}
    \int_0^T\dint{t_1}\!\!\sdots\dint{t_n}
    \\
    \times\!\Bigg(
    \!\!\Wave{w(t_1)\sdots w(t_n)}
    -(-)^\nlowest
    \!\!\Wave{w(T-t_1)\sdots w(T-t_n)}
    \!\!\Bigg)^2\\
    + \OC\left(\frac{\nu^{2\nlowest+2}}{D^{\nlowest+1}}\right)\;,
\end{multline}
where $\nlowest$ denotes the lowest order for which the squared bracket does not vanish. Inspecting \Eref{EPR_leading_order_main}, the leading order $\nlowest$ is even when $w(t)$ obeys $\SymParity$-symmetry and odd when $w(t)$ obeys $\SymTime$-symmetry.

\begin{table*}[]
    \centering
    \begin{tabular}{p{33mm}|p{25mm}|c|p{20mm}|p{20mm}|p{55mm}}
         \multicolumn{1}{c|}{$w(t)$-process} &  \multicolumn{1}{c|}{Sample graph} & \multicolumn{1}{c|}{Symmetries} & 
         \multicolumn{1}{c|}{$\EPRfull$
         } & 
         \multicolumn{1}{c|}{$\EPRpartW$} & 
         \multicolumn{1}{c}{$\EPRpartX$} \\
         \hline
            \hline
            \begin{minipage}[]{35mm}
         symmetric telegraph, \\
         $\Wave{w}=0$
         \end{minipage}& 
         \begin{minipage}{25mm}
\begin{tikzpicture}[yscale=0.7]
\draw[white] (0,0.65) -- (0,-0.65);
\draw[very thin,->] (-0.2,0)--(2.2,0) node[pos=1,above] () {$t$};
            \draw[color=white] (0,0)--(0,0.8);
\draw[very thin,->] (0,-0.6)--(0,0.7) node[pos=0.0,right,xshift=-0.5mm] () {$w$};
\draw[thick] (0,0.5) -- (0.4,0.5) -- (0.4,-0.5) -- (0.6,-0.5) -- (0.6,0.5) -- (1.2,0.5) -- (1.2,-0.5) -- (1.4,-0.5) -- (1.4,0.5) -- (1.8,0.5) -- (1.8,-0.5) -- (2,-0.5);
\end{tikzpicture}
            \end{minipage} & 
            $\SymParity$, $\SymTime$, $\SymParityTime$ & 
            \centering$\Wave{w^2}\frac{\nu^2}{D}$
            &
            \centering$0$ &
            \begin{minipage}{40mm}
            \centering$0$ 
            \end{minipage}
            \\
            \hline
            \begin{minipage}[]{35mm}asymmetric telegraph,\\
            $\Wave{w}\ne0$\end{minipage}&
            \begin{minipage}{25mm}
            \begin{tikzpicture}[yscale=0.7]
            \draw[white] (0,0.65) -- (0,-0.65);
            \draw[very thin,->] (-0.2,-0.2)--(2.2,-0.2) node[pos=1,above] () {$t$};
            \draw[color=white] (0,0)--(0,0.8);
            \draw[very thin,->] (0,-0.6)--(0,0.7) node[pos=0.0,right,xshift=-0.5mm] () {$w$};
            \draw[thick] (0,0.5) -- (0.4,0.5) -- (0.4,-0.5) -- (0.6,-0.5) -- (0.6,0.5) -- (1.2,0.5) -- (1.2,-0.5) -- (1.4,-0.5) -- (1.4,0.5) -- (1.8,0.5) -- (1.8,-0.5) -- (2,-0.5);
            \node at (2.1,0.5) {$w_+$};
            \node at (1.0,-0.5) {$w_-$};
            \end{tikzpicture}
            \end{minipage} &
            $\SymTime$, $\NotSymParity$ & 
            \centering$\Wave{w^2}\frac{\nu^2}{D}$
            &
            \centering$0$ &
            $\langle{w}\rangle_\nu^2\frac{\nu^2}{D}+\sdots$
            \hfill\Eref{asym_EPR_with_drift} \\
            \hline
         \begin{minipage}[]{35mm}
         asymmetric telegraph,\\ 
         $\Wave{w}=0$
         \end{minipage} & 
            \begin{minipage}{25mm}
            \begin{tikzpicture}[yscale=0.75]
            \draw[white] (0,0.65) -- (0,-0.65);
            \draw[very thin,->] (-0.2,-0.2)--(2.2,-0.2) node[pos=1,above] () {$t$};
            \draw[color=white] (0,0)--(0,0.8);
            \draw[very thin,->] (0,-0.6)--(0,0.7) node[pos=0.0,right,xshift=-0.9mm] () {$w$};
            \draw[thick] (0,0.5) -- (0.3,0.5) -- (0.3,-0.5) -- (0.95,-0.5) -- (0.95,0.5) -- (1.3,0.5) -- (1.3,-0.5) -- (1.7,-0.5) -- (1.7,0.5) -- (1.8,0.5) -- (1.8,-0.5) -- (2,-0.5);
            \node at (2.1,0.5) {$w_+$};
            \node at (1.2,-0.55) {$w_-$};
            \end{tikzpicture}
            \end{minipage} &
            $\SymTime$, $\NotSymParity$ & 
            \centering$\Wave{w^2}\frac{\nu^2}{D}$
            &
            \centering$0$ &
            $\frac{-1}{16}\left(\frac{\nu^2}{D}\right)^3\frac{w_+^3w_-^3}{\alpha_+\alpha_-}\left(\frac{\alpha_+-\alpha_-}{\alpha_+ + \alpha_-}\right)^2+\sdots\qquad\quad$
            \hspace*{35mm}\hfill$\phantom{}$\Eref{asym_EPR_without_drift} \\
            \hline
         \begin{minipage}[]{35mm}
         resetting,\\ $\Wave{w}=0$ 
         \end{minipage} & 
\begin{minipage}{25mm}
\begin{tikzpicture}[yscale=0.75]
            \draw[white] (0,0.65) -- (0,-0.65);
            \draw[very thin,->] (-0.2,0)--(2.2,0) node[pos=1,above] () {$t$};
            \draw[color=white] (0,0)--(0,0.8);
            \draw[very thin,->] (0,-0.6)--(0,0.7) node[pos=0.8,right,xshift=-0.5mm] () {$w$};
            \begin{scope}[yscale=1.5]
            \draw[thick] (0.000,0.000) -- (0.032,0.053) -- (0.064,0.110) -- (0.096,0.034) -- 
            (0.128,0.085) -- (0.160,0.099) -- (0.192,0.155) -- (0.224,0.191) -- (0.256,0.173) -- 
            (0.288,0.236) -- (0.320,0.208) -- (0.352,0.217) -- (0.384,0.215) -- (0.416,0.299) -- 
            (0.448,0.344) -- (0.480,0.316) -- (0.512,0.381) -- (0.544,0.361)-- 
            (0.544,0.0) -- 
            (0.576,0.000) -- (0.608,-0.051) -- (0.640,-0.003) -- (0.672,-0.012) -- (0.704,-0.072) -- (0.736,-0.045) -- (0.768,-0.092) -- (0.800,-0.184) -- 
            (0.832,-0.173) -- (0.864,-0.203) -- (0.896,-0.318) -- (0.928,-0.204) -- 
            (0.960,-0.217) -- (0.992,-0.288)-- 
            (0.992,0.0) -- 
            (1.024,0.000) -- (1.056,-0.044) -- (1.088,-0.007) -- (1.120,0.082) -- (1.152,-0.007) -- (1.184,0.025) -- (1.216,-0.063) -- (1.248,-0.063) -- (1.280,-0.087) -- (1.312,-0.067) -- (1.344,-0.082) -- (1.376,-0.164) -- (1.408,-0.143) -- (1.440,-0.137) -- (1.472,-0.105) -- (1.504,-0.184)-- 
            (1.504,-0.0) -- 
            (1.536,0.000) -- (1.568,-0.012) -- (1.600,0.040) -- (1.632,0.102) -- 
            (1.664,0.178) -- (1.696,0.099) -- (1.728,0.124) -- (1.760,0.231) -- (1.792,0.213)
-- (1.824,0.283) -- (1.856,0.268) -- (1.888,0.338) -- (1.920,0.305)
-- (1.952,0.350)-- (1.952,0.0) -- (1.984,-0.022) -- (2.000,-0.029);
            \end{scope}

\end{tikzpicture}
  \end{minipage}
	    & 
            $\SymParity$, $\NotSymTime$ & 
            \centering$\infty$ &
            \centering$\infty$ &
            $\frac{9}{32}\left(\frac{\nu^2}{D}\right)^4\frac{\diffw^4}{\resettingRate^7}+\sdots$ \hfill\Eref{resetting_epr} \\
            \hline
         \begin{minipage}[]{35mm}
         $3$-state cyclic,\\ 
         $\Wave{w}=0$ 
         \end{minipage} & 
            \begin{minipage}{25mm}
            \begin{tikzpicture}[yscale=0.75]
            \draw[white] (0,0.65) -- (0,-0.65);
            \draw[very thin,->] (-0.2,0)--(2.2,0) node[pos=1,below] () {$t$};
            \draw[color=white] (0,0)--(0,0.8);
            \draw[very thin,->] (0,-0.6)--(0,0.7) node[pos=0.0,right,xshift=-0.75mm,yshift=-0.5mm] () {$w$};
            \draw[thick] (0,0.5) -- (0.2,0.5) -- (0.2,-0.5) -- (0.4,-0.5) -- (0.4,0) -- (0.7,0) -- (0.7,0.5) -- (1.1,0.5) -- (1.1,-0.5) -- (1.4,-0.5) -- (1.4,0) -- (1.6,0) -- (1.6,0.5) -- (1.8,0.5) -- (1.8,-0.5) -- (2,-0.5);
            \node at (2.1,0.5) {$w_+$};
            \node at (0.7,-0.5) {$w_-$};
            \end{tikzpicture}
            \end{minipage} &
            $\NotSymParity$, $\NotSymTime$, $\SymParityTime$ & 
            \centering$\Wave{w^2}\frac{\nu^2}{D}+\EPRpartW$ &
            \centering$(\alpha-\beta)\ln\left(\frac{\alpha}{\beta}\right)$ &
            \centering
            $0$ 
    \end{tabular}
    \caption{Different $w(t)$-processes give rise to different trajectories $x(t)$, \Eref{langevin}, and therefore different EPR, as discussed in the main text. The first three rows produce conventional Run-and-Tumble behaviour with different parameters. The resetting noise in the fourth row  is more akin to active Brownian particles, as $w(t)$ evolves piecewise continuously in time. The last row is a variation of the first. Symmetries of the $w$-process are indicated in the third column. If broken, they are struck through.
    }
    \tlabel{processes}
\end{table*}

\textit{Applications: } In the following we consider several processes to illustrate the calculation of their partial EPR on the basis of \Erefs{partial_EP_expansion_free} and \eref{EPR_leading_order_main}. 
The sole mathematical challenge is to determine higher order correlation functions and cumulants of $w(t)$, relegated to the supplement. The results are summarised in \Tref{processes}. The first three are based on $w(t)\in\{w_+,w_-\}$ determined by a telegraph process, which switches from $w_+$ to $w_-$ with rate $\alpha_-$ and back with rate $\alpha_+$. From first principles \cite{cocconi2020entropy, paoluzziEntropyProductionRunandTumble2024a} $\EPRfull=\nu^2\Wave{w^2}/D$ with 
$\Wave{w^2}\ge\WaveS{w}{2}$ 
and $\Wave{w}=(w_+\alpha_- + w_-\alpha_+)/(\alpha_++\alpha_-)$. 
This $w(t)$-process is bound to be $\SymTime$-symmetric, even when $\alpha_+\ne\alpha_-$, as these rates merely determine occupation times in the two states. The only terms that contribute in \Eref{partial_EP_expansion_free} are therefore those with odd $n$.

Iff $w_+=-w_-$ and $\alpha_+=\alpha_-$, the process is also $\SymParity$-symmetric, so that $\EPRpartX=0$, which is expected as $x(t)$ is indistinguishable from its reverse $x(T-t)$, \Fref{process_illustrations}a. If $\Wave{w}\ne0$, then \Eref{partial_EP_expansion_free} gives
\begin{equation}\elabel{asym_EPR_with_drift}
 \EPRpartX=\frac{\nu^2}{D}\WaveS{w}{2} + \OC\left(\frac{\nu^4}{D^2}\right)\;,
\end{equation}
as would be expected from a process with net current $\nu\Wave{w}$. 
If the net drift vanishes, $\Wave{w}=0$, \Fref{process_illustrations}b, the leading order becomes, \SMref{sm_asym_tel},
\begin{equation}\elabel{asym_EPR_without_drift}
 \EPRpartX=-\frac{1}{16}\left(\frac{\nu^2}{D}\right)^3\frac{w_+^3w_-^3}{\alpha_+\alpha_-}\left(\frac{\alpha_+-\alpha_-}{\alpha_+ + \alpha_-}\right)^2+
 \OC\left(\frac{\nu^8}{D^4}\right)\ .
\end{equation}
When $\SymParity$-symmetry is restored by taking $\alpha_+=\alpha_-$ and $w_+=-w_-$, 
the right-hand sides of \Erefs{asym_EPR_with_drift} and \eref{asym_EPR_without_drift} vanish as expected.

As a $w(t)$-process that is $\SymParity$-symmetric but not $\SymTime$-symmetric, we choose the case of diffusion with diffusivity $\diffw$ and stochastic resetting to $w(t)=0$ with Poissonian rate $\resettingRate$  \cite{Switkes:2004,PhysRevLett.106.160601}, \Fref{process_illustrations}c. As there is no bias, the statistics of $w(t)$ are independent of its sign. $\SymTime$-symmetry is broken by the instantaneous resetting to the origin, causing both $\EPRfull$ and $\EPRpartW$ to diverge. As derived in \SMref{resetting_epr} using \cite{Maziya:2023},
\begin{equation}\elabel{resetting_epr}
 \EPRpartX=\frac{9}{32}\left(\frac{\nu^2}{D}\right)^4\frac{\diffw^4}{\resettingRate^7}+ 
 \OC\left( \frac{\nu^{10}}{D^5} \right) \ .
\end{equation}

We finally consider a $w(t)$-process which is neither $\SymParity$ nor $\SymTime$-symmetric, but $\SymParityTime$-symmetric, given by a cyclic change from $w_+$ to $w_-=-w_+$ to $0$ with rate $\alpha$ and backwards with rate a rate $\beta<\alpha$ \cite{power_extraction}. The EPR solely of the $w$-process is $\EPRpartW=(\alpha-\beta)\ln(\alpha/\beta)$ \cite{cocconi2020entropy}. The full EPR is, equally from first principles, $\EPRfull=\EPRpartW+\Wave{w^2}/D$ with $\Wave{w^2}=2w_+^2/3$, while the partial entropy production of the $x(t)$-process by itself vanishes, $\EPRpartX=0$.

\textit{Discussion and Outlook:}
As a rule of thumb, the harder it is to qualitatively distinguish forward and reversed spatial trajectories of a process, the higher the order of the leading term in \Eref{partial_EP_expansion_free}. Spatial trajectories of an asymmetric RnT particle with non-zero net drift can trivially be distinguished from their time-reverse due to the swapped drift direction, and have leading order partial EPR $\propto \nu^2 /D$, \ie $\nlowest=1$, \Eref{asym_EPR_with_drift} and \Tref{processes}. Trajectories of asymmetric RnT particles with zero net drift, \Fref{process_illustrations}b, differ more subtly from their time-reverse, but can be distinguished as the distribution of $\dot{x}(t)$ will be different for the forward and reversed trajectories due to the parity asymmetry of $w(t)$. This is reflected in the fact that the leading order contribution is $\nlowest=3$. In the case of stochastic resetting, \Fref{process_illustrations}c, time-irreversibility is very difficult to see with the naked eye but reveals itself through a sudden jump to zero velocity, \eg at $t=1.3, 5.1$. Such a faint effect results in the leading order contribution to the partial EPR appearing at $\nlowest=4$. 

Entropy production measures ``how broken'' $\SymTime$-symmetry is \cite{gaspard2004time,FodorETAL:2016}, \Erefs{EP_full} and \eref{partial_EP_defn}. How come the $\SymParityTime$-symmetry and not just the $\SymTime$-symmetry of $w(t)$ enters intro the entropy production of $x(t)$ as given by \Eref{langevin}? The answer is found by writing the equation of motion of $x^R(t)=x(T-t)$, which produces $\xdot^R(t)=-\xdot(T-t)=-\nu w(T-t)-\xi(T-t)$, so that the statistics of $x^R(t)$ is identical to that of $x(t)$ if the statistics of $-w(T-t)$ and $-\xi(T-t)$ are identical to that of $w(t)$ and $\xi(t)$ respectively, \ie when their statistics is $\SymParityTime$-symmetric. In other words, the reason why the $\SymParity$-symmetry  of $w(t)$ matters for the EPR of $x(t)$ is that it features on the right its time-derivative. One might think of extensions to \Eref{langevin} that make other symmetries of $w(t)$ visible in the entropy production of $x(t)$.

To use the above formalism, the mathematical challenge is to calculate correlators and higher order moments, \SMref{leading_order}. This is still much easier than calculating full path integrals, \Eref{partial_EP_defn}, but can be challenging in some cases. The formalism is readily extended to incorporate potentials \cite{KnightPruessner:2025b}, in which case even the simplest run-and-tumble setup produces entropy. The required correlators for a harmonic potential are available from \cite{garcia2021run}, but are quite unwieldy.

A thermodynamic interpretation of the partial entropy production is an important topic of future work. The level of description of a system determines how it can be manipulated \cite[][Ch.~5]{Jaynes1992}. Connection could be made between partial EPR and maximum extractable power in partially hidden systems \cite{garciamillan2024optimalclosedloopcontrolactive}. The framework could also be extended to capture the EPR of a system with an \textit{inferred} internal state, as in \cite{power_extraction, SezikKnightAlstonCocconi:2025b}.

\begin{acknowledgments}
The authors thank L Cocconi, J Fry, R Garcia-Millan, H J Jensen, S Loos and C Roberts for useful discussions. JK acknowledges support from the Engineering and Physical Sciences Research Council (grant number 2620369).
\end{acknowledgments}

\bibliography{references}

\providecommand{\noopsort}[1]{}\providecommand{\singleletter}[1]{#1}%
\begin{thebibliography}{44}%
\makeatletter
\providecommand \@ifxundefined [1]{%
 \@ifx{#1\undefined}
}%
\providecommand \@ifnum [1]{%
 \ifnum #1\expandafter \@firstoftwo
 \else \expandafter \@secondoftwo
 \fi
}%
\providecommand \@ifx [1]{%
 \ifx #1\expandafter \@firstoftwo
 \else \expandafter \@secondoftwo
 \fi
}%
\providecommand \natexlab [1]{#1}%
\providecommand \enquote  [1]{``#1''}%
\providecommand \bibnamefont  [1]{#1}%
\providecommand \bibfnamefont [1]{#1}%
\providecommand \citenamefont [1]{#1}%
\providecommand \href@noop [0]{\@secondoftwo}%
\providecommand \href [0]{\begingroup \@sanitize@url \@href}%
\providecommand \@href[1]{\@@startlink{#1}\@@href}%
\providecommand \@@href[1]{\endgroup#1\@@endlink}%
\providecommand \@sanitize@url [0]{\catcode `\\12\catcode `\$12\catcode
  `\&12\catcode `\#12\catcode `\^12\catcode `\_12\catcode `\%12\relax}%
\providecommand \@@startlink[1]{}%
\providecommand \@@endlink[0]{}%
\providecommand \url  [0]{\begingroup\@sanitize@url \@url }%
\providecommand \@url [1]{\endgroup\@href {#1}{\urlprefix }}%
\providecommand \urlprefix  [0]{URL }%
\providecommand \Eprint [0]{\href }%
\providecommand \doibase [0]{https://doi.org/}%
\providecommand \selectlanguage [0]{\@gobble}%
\providecommand \bibinfo  [0]{\@secondoftwo}%
\providecommand \bibfield  [0]{\@secondoftwo}%
\providecommand \translation [1]{[#1]}%
\providecommand \BibitemOpen [0]{}%
\providecommand \bibitemStop [0]{}%
\providecommand \bibitemNoStop [0]{.\EOS\space}%
\providecommand \EOS [0]{\spacefactor3000\relax}%
\providecommand \BibitemShut  [1]{\csname bibitem#1\endcsname}%
\let\auto@bib@innerbib\@empty
\bibitem [{\citenamefont {Montroll}\ and\ \citenamefont
  {Green}(1954)}]{montrollStatisticalMechanicsTransport1954}%
  \BibitemOpen
  \bibfield  {author} {\bibinfo {author} {\bibfnamefont {E.~W.}\ \bibnamefont
  {Montroll}}\ and\ \bibinfo {author} {\bibfnamefont {M.~S.}\ \bibnamefont
  {Green}},\ }\bibfield  {title} {\bibinfo {title} {Statistical {{Mechanics}}
  of {{Transport}} and {{Nonequilibrium Processes}}},\ }\href
  {https://doi.org/10.1146/annurev.pc.05.100154.002313} {\bibfield  {journal}
  {\bibinfo  {journal} {Annu. Rev. Phys. Chem.}\ }\textbf {\bibinfo {volume}
  {5}},\ \bibinfo {pages} {449} (\bibinfo {year} {1954})}\BibitemShut {NoStop}%
\bibitem [{\citenamefont {Astumian}\ and\ \citenamefont
  {Hänggi}(2002)}]{astumianBrownianMotors2002}%
  \BibitemOpen
  \bibfield  {author} {\bibinfo {author} {\bibfnamefont {R.~D.}\ \bibnamefont
  {Astumian}}\ and\ \bibinfo {author} {\bibfnamefont {P.}~\bibnamefont
  {Hänggi}},\ }\bibfield  {title} {\bibinfo {title} {Brownian {{Motors}}},\
  }\href {https://doi.org/10.1063/1.1535005} {\bibfield  {journal} {\bibinfo
  {journal} {Phys. Today}\ }\textbf {\bibinfo {volume} {55}},\ \bibinfo {pages}
  {33} (\bibinfo {year} {2002})}\BibitemShut {NoStop}%
\bibitem [{\citenamefont
  {Michaelian}(2011)}]{michaelianEntropyProductionOrigin2011}%
  \BibitemOpen
  \bibfield  {author} {\bibinfo {author} {\bibfnamefont {K.}~\bibnamefont
  {Michaelian}},\ }\bibfield  {title} {\bibinfo {title} {Entropy {{Production}}
  and the {{Origin}} of {{Life}}},\ }\href
  {https://doi.org/10.4236/jmp.2011.226069} {\bibfield  {journal} {\bibinfo
  {journal} {J. Mod. Phys.}\ }\textbf {\bibinfo {volume} {2}},\ \bibinfo
  {pages} {595} (\bibinfo {year} {2011})}\BibitemShut {NoStop}%
\bibitem [{\citenamefont {Thurley}\ \emph {et~al.}(2014)\citenamefont
  {Thurley}, \citenamefont {Tovey}, \citenamefont {Moenke}, \citenamefont
  {Prince}, \citenamefont {Meena}, \citenamefont {Thomas}, \citenamefont
  {Skupin}, \citenamefont {Taylor},\ and\ \citenamefont
  {Falcke}}]{thurleyReliableEncodingStimulus2014}%
  \BibitemOpen
  \bibfield  {author} {\bibinfo {author} {\bibfnamefont {K.}~\bibnamefont
  {Thurley}}, \bibinfo {author} {\bibfnamefont {S.~C.}\ \bibnamefont {Tovey}},
  \bibinfo {author} {\bibfnamefont {G.}~\bibnamefont {Moenke}}, \bibinfo
  {author} {\bibfnamefont {V.~L.}\ \bibnamefont {Prince}}, \bibinfo {author}
  {\bibfnamefont {A.}~\bibnamefont {Meena}}, \bibinfo {author} {\bibfnamefont
  {A.~P.}\ \bibnamefont {Thomas}}, \bibinfo {author} {\bibfnamefont
  {A.}~\bibnamefont {Skupin}}, \bibinfo {author} {\bibfnamefont {C.~W.}\
  \bibnamefont {Taylor}},\ and\ \bibinfo {author} {\bibfnamefont
  {M.}~\bibnamefont {Falcke}},\ }\bibfield  {title} {\bibinfo {title} {Reliable
  {{Encoding}} of {{Stimulus Intensities Within Random Sequences}} of
  {{Intracellular Ca2}}+ {{Spikes}}},\ }\href
  {https://doi.org/10.1126/scisignal.2005237} {\bibfield  {journal} {\bibinfo
  {journal} {Science Signaling}\ }\textbf {\bibinfo {volume} {7}},\ \bibinfo
  {pages} {ra59} (\bibinfo {year} {2014})}\BibitemShut {NoStop}%
\bibitem [{\citenamefont {Lan}\ \emph {et~al.}(2012)\citenamefont {Lan},
  \citenamefont {Sartori}, \citenamefont {Neumann}, \citenamefont {Sourjik},\
  and\ \citenamefont {Tu}}]{lanEnergySpeedAccuracy2012}%
  \BibitemOpen
  \bibfield  {author} {\bibinfo {author} {\bibfnamefont {G.}~\bibnamefont
  {Lan}}, \bibinfo {author} {\bibfnamefont {P.}~\bibnamefont {Sartori}},
  \bibinfo {author} {\bibfnamefont {S.}~\bibnamefont {Neumann}}, \bibinfo
  {author} {\bibfnamefont {V.}~\bibnamefont {Sourjik}},\ and\ \bibinfo {author}
  {\bibfnamefont {Y.}~\bibnamefont {Tu}},\ }\bibfield  {title} {\bibinfo
  {title} {The energy--speed--accuracy trade-off in sensory adaptation},\
  }\href {https://doi.org/10.1038/nphys2276} {\bibfield  {journal} {\bibinfo
  {journal} {Nature Physics}\ }\textbf {\bibinfo {volume} {8}},\ \bibinfo
  {pages} {422} (\bibinfo {year} {2012})}\BibitemShut {NoStop}%
\bibitem [{\citenamefont {Harvey}\ \emph {et~al.}(2023)\citenamefont {Harvey},
  \citenamefont {Lahiri},\ and\ \citenamefont
  {Ganguli}}]{harveyUniversalEnergyaccuracyTradeoffs2023}%
  \BibitemOpen
  \bibfield  {author} {\bibinfo {author} {\bibfnamefont {S.~E.}\ \bibnamefont
  {Harvey}}, \bibinfo {author} {\bibfnamefont {S.}~\bibnamefont {Lahiri}},\
  and\ \bibinfo {author} {\bibfnamefont {S.}~\bibnamefont {Ganguli}},\
  }\bibfield  {title} {\bibinfo {title} {Universal energy-accuracy tradeoffs in
  nonequilibrium cellular sensing},\ }\href
  {https://doi.org/10.1103/PhysRevE.108.014403} {\bibfield  {journal} {\bibinfo
   {journal} {Physical Review E}\ }\textbf {\bibinfo {volume} {108}},\ \bibinfo
  {pages} {014403} (\bibinfo {year} {2023})}\BibitemShut {NoStop}%
\bibitem [{\citenamefont
  {England}(2013)}]{englandStatisticalPhysicsSelfreplication2013}%
  \BibitemOpen
  \bibfield  {author} {\bibinfo {author} {\bibfnamefont {J.~L.}\ \bibnamefont
  {England}},\ }\bibfield  {title} {\bibinfo {title} {Statistical physics of
  self-replication},\ }\href {https://doi.org/10.1063/1.4818538} {\bibfield
  {journal} {\bibinfo  {journal} {The Journal of Chemical Physics}\ }\textbf
  {\bibinfo {volume} {139}},\ \bibinfo {pages} {121923} (\bibinfo {year}
  {2013})}\BibitemShut {NoStop}%
\bibitem [{\citenamefont {Dieball}\ and\ \citenamefont
  {Godec}(2025)}]{dieballPerspectiveTimeIrreversibility2025}%
  \BibitemOpen
  \bibfield  {author} {\bibinfo {author} {\bibfnamefont {C.}~\bibnamefont
  {Dieball}}\ and\ \bibinfo {author} {\bibfnamefont {A.}~\bibnamefont
  {Godec}},\ }\bibfield  {title} {\bibinfo {title} {Perspective: {{Time}}
  irreversibility in systems observed at coarse resolution},\ }\href
  {https://doi.org/10.1063/5.0251089} {\bibfield  {journal} {\bibinfo
  {journal} {J. Chem. Phys.}\ }\textbf {\bibinfo {volume} {162}},\ \bibinfo
  {pages} {090901} (\bibinfo {year} {2025})}\BibitemShut {NoStop}%
\bibitem [{\citenamefont {Fodor}\ \emph {et~al.}(2016)\citenamefont {Fodor},
  \citenamefont {Nardini}, \citenamefont {Cates}, \citenamefont {Tailleur},
  \citenamefont {Visco},\ and\ \citenamefont {van Wijland}}]{FodorETAL:2016}%
  \BibitemOpen
  \bibfield  {author} {\bibinfo {author} {\bibfnamefont {E.}~\bibnamefont
  {Fodor}}, \bibinfo {author} {\bibfnamefont {C.}~\bibnamefont {Nardini}},
  \bibinfo {author} {\bibfnamefont {M.~E.}\ \bibnamefont {Cates}}, \bibinfo
  {author} {\bibfnamefont {J.}~\bibnamefont {Tailleur}}, \bibinfo {author}
  {\bibfnamefont {P.}~\bibnamefont {Visco}},\ and\ \bibinfo {author}
  {\bibfnamefont {F.}~\bibnamefont {van Wijland}},\ }\bibfield  {title}
  {\bibinfo {title} {How far from equilibrium is active matter?},\ }\href
  {https://doi.org/10.1103/PhysRevLett.117.038103} {\bibfield  {journal}
  {\bibinfo  {journal} {Phys. Rev. Lett.}\ }\textbf {\bibinfo {volume} {117}},\
  \bibinfo {pages} {038103} (\bibinfo {year} {2016})}\BibitemShut {NoStop}%
\bibitem [{\citenamefont {Bisker}\ \emph {et~al.}(2017)\citenamefont {Bisker},
  \citenamefont {Polettini}, \citenamefont {Gingrich},\ and\ \citenamefont
  {Horowitz}}]{biskerHierarchicalBoundsEntropy2017}%
  \BibitemOpen
  \bibfield  {author} {\bibinfo {author} {\bibfnamefont {G.}~\bibnamefont
  {Bisker}}, \bibinfo {author} {\bibfnamefont {M.}~\bibnamefont {Polettini}},
  \bibinfo {author} {\bibfnamefont {T.~R.}\ \bibnamefont {Gingrich}},\ and\
  \bibinfo {author} {\bibfnamefont {J.~M.}\ \bibnamefont {Horowitz}},\
  }\bibfield  {title} {\bibinfo {title} {Hierarchical bounds on entropy
  production inferred from partial information},\ }\href
  {https://doi.org/10.1088/1742-5468/aa8c0d} {\bibfield  {journal} {\bibinfo
  {journal} {J. Stat. Mech.}\ }\textbf {\bibinfo {volume} {2017}},\ \bibinfo
  {pages} {093210} (\bibinfo {year} {2017})}\BibitemShut {NoStop}%
\bibitem [{\citenamefont {Nitzan}\ \emph {et~al.}(2023)\citenamefont {Nitzan},
  \citenamefont {Ghosal},\ and\ \citenamefont
  {Bisker}}]{nitzanUniversalBoundsEntropy2023}%
  \BibitemOpen
  \bibfield  {author} {\bibinfo {author} {\bibfnamefont {E.}~\bibnamefont
  {Nitzan}}, \bibinfo {author} {\bibfnamefont {A.}~\bibnamefont {Ghosal}},\
  and\ \bibinfo {author} {\bibfnamefont {G.}~\bibnamefont {Bisker}},\
  }\bibfield  {title} {\bibinfo {title} {Universal bounds on entropy production
  inferred from observed statistics},\ }\href
  {https://doi.org/10.1103/PhysRevResearch.5.043251} {\bibfield  {journal}
  {\bibinfo  {journal} {Phys. Rev. Res.}\ }\textbf {\bibinfo {volume} {5}},\
  \bibinfo {pages} {043251} (\bibinfo {year} {2023})}\BibitemShut {NoStop}%
\bibitem [{\citenamefont {Mart{\'i}nez}\ \emph {et~al.}(2019)\citenamefont
  {Mart{\'i}nez}, \citenamefont {Bisker}, \citenamefont {Horowitz},\ and\
  \citenamefont {Parrondo}}]{martinezInferringBrokenDetailed2019}%
  \BibitemOpen
  \bibfield  {author} {\bibinfo {author} {\bibfnamefont {I.~A.}\ \bibnamefont
  {Mart{\'i}nez}}, \bibinfo {author} {\bibfnamefont {G.}~\bibnamefont
  {Bisker}}, \bibinfo {author} {\bibfnamefont {J.~M.}\ \bibnamefont
  {Horowitz}},\ and\ \bibinfo {author} {\bibfnamefont {J.~M.~R.}\ \bibnamefont
  {Parrondo}},\ }\bibfield  {title} {\bibinfo {title} {Inferring broken
  detailed balance in the absence of observable currents},\ }\href
  {https://doi.org/10.1038/s41467-019-11051-w} {\bibfield  {journal} {\bibinfo
  {journal} {Nat. Commun.}\ }\textbf {\bibinfo {volume} {10}},\ \bibinfo
  {pages} {3542} (\bibinfo {year} {2019})}\BibitemShut {NoStop}%
\bibitem [{\citenamefont {Van Der~Meer}\ \emph {et~al.}(2022)\citenamefont {Van
  Der~Meer}, \citenamefont {Ertel},\ and\ \citenamefont
  {Seifert}}]{vandermeerThermodynamicInferencePartially2022}%
  \BibitemOpen
  \bibfield  {author} {\bibinfo {author} {\bibfnamefont {J.}~\bibnamefont {Van
  Der~Meer}}, \bibinfo {author} {\bibfnamefont {B.}~\bibnamefont {Ertel}},\
  and\ \bibinfo {author} {\bibfnamefont {U.}~\bibnamefont {Seifert}},\
  }\bibfield  {title} {\bibinfo {title} {Thermodynamic {{Inference}} in
  {{Partially Accessible Markov Networks}}: {{A Unifying Perspective}} from
  {{Transition-Based Waiting Time Distributions}}},\ }\href
  {https://doi.org/10.1103/PhysRevX.12.031025} {\bibfield  {journal} {\bibinfo
  {journal} {Phys. Rev. X}\ }\textbf {\bibinfo {volume} {12}},\ \bibinfo
  {pages} {031025} (\bibinfo {year} {2022})}\BibitemShut {NoStop}%
\bibitem [{\citenamefont {Ertel}\ \emph {et~al.}(2022)\citenamefont {Ertel},
  \citenamefont {{van der Meer}},\ and\ \citenamefont
  {Seifert}}]{ertelOperationallyAccessibleUncertainty2022}%
  \BibitemOpen
  \bibfield  {author} {\bibinfo {author} {\bibfnamefont {B.}~\bibnamefont
  {Ertel}}, \bibinfo {author} {\bibfnamefont {J.}~\bibnamefont {{van der
  Meer}}},\ and\ \bibinfo {author} {\bibfnamefont {U.}~\bibnamefont
  {Seifert}},\ }\bibfield  {title} {\bibinfo {title} {Operationally accessible
  uncertainty relations for thermodynamically consistent semi-{{Markov}}
  processes},\ }\href {https://doi.org/10.1103/PhysRevE.105.044113} {\bibfield
  {journal} {\bibinfo  {journal} {Phys. Rev. E}\ }\textbf {\bibinfo {volume}
  {105}},\ \bibinfo {pages} {044113} (\bibinfo {year} {2022})}\BibitemShut
  {NoStop}%
\bibitem [{\citenamefont {Deg{\"u}nther}\ \emph {et~al.}(2024)\citenamefont
  {Deg{\"u}nther}, \citenamefont {Van Der~Meer},\ and\ \citenamefont
  {Seifert}}]{deguntherFluctuatingEntropyProduction2024}%
  \BibitemOpen
  \bibfield  {author} {\bibinfo {author} {\bibfnamefont {J.}~\bibnamefont
  {Deg{\"u}nther}}, \bibinfo {author} {\bibfnamefont {J.}~\bibnamefont {Van
  Der~Meer}},\ and\ \bibinfo {author} {\bibfnamefont {U.}~\bibnamefont
  {Seifert}},\ }\bibfield  {title} {\bibinfo {title} {Fluctuating entropy
  production on the coarse-grained level: {{Inference}} and localization of
  irreversibility},\ }\href {https://doi.org/10.1103/PhysRevResearch.6.023175}
  {\bibfield  {journal} {\bibinfo  {journal} {Phys. Rev. Res.}\ }\textbf
  {\bibinfo {volume} {6}},\ \bibinfo {pages} {023175} (\bibinfo {year}
  {2024})}\BibitemShut {NoStop}%
\bibitem [{\citenamefont {Degünther}\ \emph {et~al.}(2024)\citenamefont
  {Degünther}, \citenamefont {van~der Meer},\ and\ \citenamefont
  {Seifert}}]{degunthervandermeerseifertcalmodulin}%
  \BibitemOpen
  \bibfield  {author} {\bibinfo {author} {\bibfnamefont {J.}~\bibnamefont
  {Degünther}}, \bibinfo {author} {\bibfnamefont {J.}~\bibnamefont {van~der
  Meer}},\ and\ \bibinfo {author} {\bibfnamefont {U.}~\bibnamefont {Seifert}},\
  }\bibfield  {title} {\bibinfo {title} {General theory for localizing the
  where and when of entropy production meets single-molecule experiments},\
  }\href {https://doi.org/10.1073/pnas.2405371121} {\bibfield  {journal}
  {\bibinfo  {journal} {Proceedings of the National Academy of Sciences}\
  }\textbf {\bibinfo {volume} {121}},\ \bibinfo {pages} {e2405371121} (\bibinfo
  {year} {2024})},\ \Eprint
  {https://arxiv.org/abs/https://www.pnas.org/doi/pdf/10.1073/pnas.2405371121}
  {https://www.pnas.org/doi/pdf/10.1073/pnas.2405371121} \BibitemShut {NoStop}%
\bibitem [{\citenamefont {Blom}\ \emph {et~al.}(2024)\citenamefont {Blom},
  \citenamefont {Song}, \citenamefont {Vouga}, \citenamefont {Godec},\ and\
  \citenamefont {Makarov}}]{blomMilestoningEstimatorsDissipation2024}%
  \BibitemOpen
  \bibfield  {author} {\bibinfo {author} {\bibfnamefont {K.}~\bibnamefont
  {Blom}}, \bibinfo {author} {\bibfnamefont {K.}~\bibnamefont {Song}}, \bibinfo
  {author} {\bibfnamefont {E.}~\bibnamefont {Vouga}}, \bibinfo {author}
  {\bibfnamefont {A.}~\bibnamefont {Godec}},\ and\ \bibinfo {author}
  {\bibfnamefont {D.~E.}\ \bibnamefont {Makarov}},\ }\bibfield  {title}
  {\bibinfo {title} {Milestoning estimators of dissipation in systems observed
  at a coarse resolution},\ }\href {https://doi.org/10.1073/pnas.2318333121}
  {\bibfield  {journal} {\bibinfo  {journal} {Proc. Natl. Acad. Sci. USA}\
  }\textbf {\bibinfo {volume} {121}},\ \bibinfo {pages} {e2318333121} (\bibinfo
  {year} {2024})}\BibitemShut {NoStop}%
\bibitem [{\citenamefont {Hartich}\ and\ \citenamefont
  {Godec}(2023)}]{hartichViolationLocalDetailed2023}%
  \BibitemOpen
  \bibfield  {author} {\bibinfo {author} {\bibfnamefont {D.}~\bibnamefont
  {Hartich}}\ and\ \bibinfo {author} {\bibfnamefont {A.}~\bibnamefont
  {Godec}},\ }\bibfield  {title} {\bibinfo {title} {Violation of local detailed
  balance upon lumping despite a clear timescale separation},\ }\href
  {https://doi.org/10.1103/PhysRevResearch.5.L032017} {\bibfield  {journal}
  {\bibinfo  {journal} {Phys. Rev. Res.}\ }\textbf {\bibinfo {volume} {5}},\
  \bibinfo {pages} {L032017} (\bibinfo {year} {2023})}\BibitemShut {NoStop}%
\bibitem [{\citenamefont {Dabelow}\ \emph {et~al.}(2019)\citenamefont
  {Dabelow}, \citenamefont {Bo},\ and\ \citenamefont
  {Eichhorn}}]{dabelowIrreversibilityActiveMatter2019}%
  \BibitemOpen
  \bibfield  {author} {\bibinfo {author} {\bibfnamefont {L.}~\bibnamefont
  {Dabelow}}, \bibinfo {author} {\bibfnamefont {S.}~\bibnamefont {Bo}},\ and\
  \bibinfo {author} {\bibfnamefont {R.}~\bibnamefont {Eichhorn}},\ }\bibfield
  {title} {\bibinfo {title} {Irreversibility in {{Active Matter Systems}}:
  {{Fluctuation Theorem}} and {{Mutual Information}}},\ }\href
  {https://doi.org/10.1103/PhysRevX.9.021009} {\bibfield  {journal} {\bibinfo
  {journal} {Phys. Rev. X}\ }\textbf {\bibinfo {volume} {9}},\ \bibinfo {pages}
  {021009} (\bibinfo {year} {2019})}\BibitemShut {NoStop}%
\bibitem [{\citenamefont {Caprini}\ \emph {et~al.}(2019)\citenamefont
  {Caprini}, \citenamefont {Marconi}, \citenamefont {Puglisi},\ and\
  \citenamefont {Vulpiani}}]{capriniEntropyProductionOrnstein2019}%
  \BibitemOpen
  \bibfield  {author} {\bibinfo {author} {\bibfnamefont {L.}~\bibnamefont
  {Caprini}}, \bibinfo {author} {\bibfnamefont {U.~M.~B.}\ \bibnamefont
  {Marconi}}, \bibinfo {author} {\bibfnamefont {A.}~\bibnamefont {Puglisi}},\
  and\ \bibinfo {author} {\bibfnamefont {A.}~\bibnamefont {Vulpiani}},\
  }\bibfield  {title} {\bibinfo {title} {The entropy production of
  {{Ornstein}}--{{Uhlenbeck}} active particles: A path integral method for
  correlations},\ }\href {https://doi.org/10.1088/1742-5468/ab14dd} {\bibfield
  {journal} {\bibinfo  {journal} {J. Stat. Mech.}\ }\textbf {\bibinfo {volume}
  {2019}},\ \bibinfo {pages} {053203} (\bibinfo {year} {2019})}\BibitemShut
  {NoStop}%
\bibitem [{\citenamefont {Cocconi}\ \emph {et~al.}(2023)\citenamefont
  {Cocconi}, \citenamefont {Knight},\ and\ \citenamefont
  {Roberts}}]{power_extraction}%
  \BibitemOpen
  \bibfield  {author} {\bibinfo {author} {\bibfnamefont {L.}~\bibnamefont
  {Cocconi}}, \bibinfo {author} {\bibfnamefont {J.}~\bibnamefont {Knight}},\
  and\ \bibinfo {author} {\bibfnamefont {C.}~\bibnamefont {Roberts}},\
  }\bibfield  {title} {\bibinfo {title} {Optimal power extraction from active
  particles with hidden states},\ }\href
  {https://doi.org/10.1103/PhysRevLett.131.188301} {\bibfield  {journal}
  {\bibinfo  {journal} {Phys. Rev. Lett.}\ }\textbf {\bibinfo {volume} {131}},\
  \bibinfo {pages} {188301} (\bibinfo {year} {2023})}\BibitemShut {NoStop}%
\bibitem [{\citenamefont {Dabelow}\ \emph {et~al.}(2021)\citenamefont
  {Dabelow}, \citenamefont {Bo},\ and\ \citenamefont
  {Eichhorn}}]{DabelowBoEichhorn:2021}%
  \BibitemOpen
  \bibfield  {author} {\bibinfo {author} {\bibfnamefont {L.}~\bibnamefont
  {Dabelow}}, \bibinfo {author} {\bibfnamefont {S.}~\bibnamefont {Bo}},\ and\
  \bibinfo {author} {\bibfnamefont {R.}~\bibnamefont {Eichhorn}},\ }\bibfield
  {title} {\bibinfo {title} {How irreversible are steady-state trajectories of
  a trapped active particle?},\ }\href
  {https://doi.org/10.1088/1742-5468/abe6fd} {\bibfield  {journal} {\bibinfo
  {journal} {J. Stat. Mech.}\ }\textbf {\bibinfo {volume} {2021}},\ \bibinfo
  {pages} {033216} (\bibinfo {year} {2021})}\BibitemShut {NoStop}%
\bibitem [{\citenamefont {Tailleur}\ and\ \citenamefont
  {Cates}(2008)}]{TailleurCates:2008}%
  \BibitemOpen
  \bibfield  {author} {\bibinfo {author} {\bibfnamefont {J.}~\bibnamefont
  {Tailleur}}\ and\ \bibinfo {author} {\bibfnamefont {M.~E.}\ \bibnamefont
  {Cates}},\ }\bibfield  {title} {\bibinfo {title} {Statistical mechanics of
  interacting run-and-tumble bacteria},\ }\href
  {https://doi.org/10.1103/PhysRevLett.100.218103} {\bibfield  {journal}
  {\bibinfo  {journal} {Phys. Rev. Lett.}\ }\textbf {\bibinfo {volume} {100}},\
  \bibinfo {pages} {218103} (\bibinfo {year} {2008})}\BibitemShut {NoStop}%
\bibitem [{\citenamefont {Cates}\ and\ \citenamefont
  {Tailleur}(2013)}]{CatesTailleur:2013}%
  \BibitemOpen
  \bibfield  {author} {\bibinfo {author} {\bibfnamefont {M.~E.}\ \bibnamefont
  {Cates}}\ and\ \bibinfo {author} {\bibfnamefont {J.}~\bibnamefont
  {Tailleur}},\ }\bibfield  {title} {\bibinfo {title} {When are active brownian
  particles and run-and-tumble particles equivalent? consequences for
  motility-induced phase separation},\ }\href
  {https://doi.org/10.1209/0295-5075/101/20010} {\bibfield  {journal} {\bibinfo
   {journal} {EPL}\ }\textbf {\bibinfo {volume} {101}},\ \bibinfo {pages}
  {20010} (\bibinfo {year} {2013})}\BibitemShut {NoStop}%
\bibitem [{\citenamefont {Gaspard}(2004)}]{gaspard2004time}%
  \BibitemOpen
  \bibfield  {author} {\bibinfo {author} {\bibfnamefont {P.}~\bibnamefont
  {Gaspard}},\ }\bibfield  {title} {\bibinfo {title} {Time-reversed dynamical
  entropy and irreversibility in {M}arkovian random processes},\ }\href
  {https://link.springer.com/article/10.1007/s10955-004-3455-1} {\bibfield
  {journal} {\bibinfo  {journal} {J. Stat. Phys.}\ }\textbf {\bibinfo {volume}
  {117}},\ \bibinfo {pages} {599} (\bibinfo {year} {2004})}\BibitemShut
  {NoStop}%
\bibitem [{\citenamefont {Cocconi}\ \emph {et~al.}(2020)\citenamefont
  {Cocconi}, \citenamefont {Garcia-Millan}, \citenamefont {Zhen}, \citenamefont
  {Buturca},\ and\ \citenamefont {Pruessner}}]{cocconi2020entropy}%
  \BibitemOpen
  \bibfield  {author} {\bibinfo {author} {\bibfnamefont {L.}~\bibnamefont
  {Cocconi}}, \bibinfo {author} {\bibfnamefont {R.}~\bibnamefont
  {Garcia-Millan}}, \bibinfo {author} {\bibfnamefont {Z.}~\bibnamefont {Zhen}},
  \bibinfo {author} {\bibfnamefont {B.}~\bibnamefont {Buturca}},\ and\ \bibinfo
  {author} {\bibfnamefont {G.}~\bibnamefont {Pruessner}},\ }\bibfield  {title}
  {\bibinfo {title} {Entropy production in exactly solvable systems},\ }\href
  {https://www.mdpi.com/1099-4300/22/11/1252} {\bibfield  {journal} {\bibinfo
  {journal} {Entropy}\ }\textbf {\bibinfo {volume} {22}},\ \bibinfo {pages}
  {1252} (\bibinfo {year} {2020})}\BibitemShut {NoStop}%
\bibitem [{\citenamefont {Sekimoto}(2012)}]{Sekimoto:2010}%
  \BibitemOpen
  \bibfield  {author} {\bibinfo {author} {\bibfnamefont {K.}~\bibnamefont
  {Sekimoto}},\ }\bibfield  {title} {\bibinfo {title} {Stochastic energetics}\
  }(\bibinfo  {publisher} {Springer-Verlag},\ \bibinfo {address} {Berlin,
  Germany},\ \bibinfo {year} {2012})\ pp.\ \bibinfo {pages}
  {XVIII,322}\BibitemShut {NoStop}%
\bibitem [{\citenamefont {T{\"a}uber}(2014)}]{Taeuber:2014}%
  \BibitemOpen
  \bibfield  {author} {\bibinfo {author} {\bibfnamefont {U.~C.}\ \bibnamefont
  {T{\"a}uber}},\ }\href@noop {} {\emph {\bibinfo {title} {Critical
  dynamics}}}\ (\bibinfo  {publisher} {Cambridge University Press},\ \bibinfo
  {address} {Cambridge, UK},\ \bibinfo {year} {2014})\ pp.\ \bibinfo {pages}
  {i--xvi,1--511}\BibitemShut {NoStop}%
\bibitem [{\citenamefont {Rold{\'a}n}\ \emph {et~al.}(2023)\citenamefont
  {Rold{\'a}n}, \citenamefont {Neri}, \citenamefont {Chetrite}, \citenamefont
  {Gupta}, \citenamefont {Pigolotti}, \citenamefont {J{\"u}licher},\ and\
  \citenamefont {Sekimoto}}]{roldanMartingalesPhysicistsTreatise2023}%
  \BibitemOpen
  \bibfield  {author} {\bibinfo {author} {\bibfnamefont {{\'E}.}~\bibnamefont
  {Rold{\'a}n}}, \bibinfo {author} {\bibfnamefont {I.}~\bibnamefont {Neri}},
  \bibinfo {author} {\bibfnamefont {R.}~\bibnamefont {Chetrite}}, \bibinfo
  {author} {\bibfnamefont {S.}~\bibnamefont {Gupta}}, \bibinfo {author}
  {\bibfnamefont {S.}~\bibnamefont {Pigolotti}}, \bibinfo {author}
  {\bibfnamefont {F.}~\bibnamefont {J{\"u}licher}},\ and\ \bibinfo {author}
  {\bibfnamefont {K.}~\bibnamefont {Sekimoto}},\ }\bibfield  {title} {\bibinfo
  {title} {Martingales for physicists: A treatise on stochastic thermodynamics
  and beyond},\ }\href {https://doi.org/10.1080/00018732.2024.2317494}
  {\bibfield  {journal} {\bibinfo  {journal} {Adv. Phys.}\ }\textbf {\bibinfo
  {volume} {72}},\ \bibinfo {pages} {1} (\bibinfo {year} {2023})}\BibitemShut
  {NoStop}%
\bibitem [{\citenamefont {Pruessner}\ and\ \citenamefont
  {Garcia-Millan}(2022)}]{PruessnerGarcia-Millan:2022}%
  \BibitemOpen
  \bibfield  {author} {\bibinfo {author} {\bibfnamefont {G.}~\bibnamefont
  {Pruessner}}\ and\ \bibinfo {author} {\bibfnamefont {R.}~\bibnamefont
  {Garcia-Millan}},\ }\bibfield  {title} {\bibinfo {title} {Field theories of
  active particle systems and their entropy production},\ }\Eprint
  {https://arxiv.org/abs/arXiv:2211.11906} {arXiv:2211.11906}  (\bibinfo {year}
  {2022}),\ \bibinfo {note} {preprint}\BibitemShut {NoStop}%
\bibitem [{\citenamefont {Pruessner}\ and\ \citenamefont
  {Knight}(2025)}]{KnightPruessner:2025b}%
  \BibitemOpen
  \bibfield  {author} {\bibinfo {author} {\bibfnamefont {G.}~\bibnamefont
  {Pruessner}}\ and\ \bibinfo {author} {\bibfnamefont {J.}~\bibnamefont
  {Knight}},\ }\bibfield  {title} {\bibinfo {title} {Entropy production of
  active particles with hidden state in potentials}} (\bibinfo {year} {2025}),\
  \bibinfo {note} {in preparation}\BibitemShut {NoStop}%
\bibitem [{\citenamefont {Paoluzzi}\ \emph {et~al.}(2024)\citenamefont
  {Paoluzzi}, \citenamefont {Puglisi},\ and\ \citenamefont
  {Angelani}}]{paoluzziEntropyProductionRunandTumble2024a}%
  \BibitemOpen
  \bibfield  {author} {\bibinfo {author} {\bibfnamefont {M.}~\bibnamefont
  {Paoluzzi}}, \bibinfo {author} {\bibfnamefont {A.}~\bibnamefont {Puglisi}},\
  and\ \bibinfo {author} {\bibfnamefont {L.}~\bibnamefont {Angelani}},\
  }\bibfield  {title} {\bibinfo {title} {Entropy {{Production}} of
  {{Run-and-Tumble Particles}}},\ }\href {https://doi.org/10.3390/e26060443}
  {\bibfield  {journal} {\bibinfo  {journal} {Entropy}\ }\textbf {\bibinfo
  {volume} {26}},\ \bibinfo {pages} {443} (\bibinfo {year} {2024})}\BibitemShut
  {NoStop}%
\bibitem [{\citenamefont {Switkes}(2004)}]{Switkes:2004}%
  \BibitemOpen
  \bibfield  {author} {\bibinfo {author} {\bibfnamefont {J.}~\bibnamefont
  {Switkes}},\ }\bibfield  {title} {\bibinfo {title} {An unbiased random walk
  with catastrophe},\ }\href
  {https://drive.google.com/file/d/1FkP-ODFN9wwPQeMw2dqk08Uz3uYslkv9/view}
  {\bibfield  {journal} {\bibinfo  {journal} {Math. Scientist}\ }\textbf
  {\bibinfo {volume} {29}},\ \bibinfo {pages} {115} (\bibinfo {year}
  {2004})}\BibitemShut {NoStop}%
\bibitem [{\citenamefont {Evans}\ and\ \citenamefont
  {Majumdar}(2011)}]{PhysRevLett.106.160601}%
  \BibitemOpen
  \bibfield  {author} {\bibinfo {author} {\bibfnamefont {M.~R.}\ \bibnamefont
  {Evans}}\ and\ \bibinfo {author} {\bibfnamefont {S.~N.}\ \bibnamefont
  {Majumdar}},\ }\bibfield  {title} {\bibinfo {title} {Diffusion with
  stochastic resetting},\ }\href
  {https://doi.org/10.1103/PhysRevLett.106.160601} {\bibfield  {journal}
  {\bibinfo  {journal} {Phys. Rev. Lett.}\ }\textbf {\bibinfo {volume} {106}},\
  \bibinfo {pages} {160601} (\bibinfo {year} {2011})}\BibitemShut {NoStop}%
\bibitem [{\citenamefont {Maziya}(2023)}]{Maziya:2023}%
  \BibitemOpen
  \bibfield  {author} {\bibinfo {author} {\bibfnamefont {M.~G.}\ \bibnamefont
  {Maziya}},\ }\emph {\bibinfo {title} {The coupon collector’s problem, the
  cover time problem, and the first passage time problem}},\ \href
  {https://doi.org/https://doi.org/10.25560/105557} {Ph.D. thesis},\ \bibinfo
  {school} {Imperial College London} (\bibinfo {year} {2023})\BibitemShut
  {NoStop}%
\bibitem [{\citenamefont {Garcia-Millan}\ and\ \citenamefont
  {Pruessner}(2021)}]{garcia2021run}%
  \BibitemOpen
  \bibfield  {author} {\bibinfo {author} {\bibfnamefont {R.}~\bibnamefont
  {Garcia-Millan}}\ and\ \bibinfo {author} {\bibfnamefont {G.}~\bibnamefont
  {Pruessner}},\ }\bibfield  {title} {\bibinfo {title} {Run-and-tumble motion
  in a harmonic potential: {F}ield theory and entropy production},\ }\href@noop
  {} {\bibfield  {journal} {\bibinfo  {journal} {J. Stat. Mech.}\ }\textbf
  {\bibinfo {volume} {2021}},\ \bibinfo {pages} {063203} (\bibinfo {year}
  {2021})}\BibitemShut {NoStop}%
\bibitem [{\citenamefont {Jaynes}(1992)}]{Jaynes1992}%
  \BibitemOpen
  \bibfield  {author} {\bibinfo {author} {\bibfnamefont {E.~T.}\ \bibnamefont
  {Jaynes}},\ }\bibfield  {title} {\bibinfo {title} {The gibbs paradox},\ }in\
  \href {https://doi.org/10.1007/978-94-017-2219-3_1} {\emph {\bibinfo
  {booktitle} {Maximum Entropy and Bayesian Methods: {{Seattle}}, 1991}}},\
  \bibinfo {editor} {edited by\ \bibinfo {editor} {\bibfnamefont {C.~R.}\
  \bibnamefont {Smith}}, \bibinfo {editor} {\bibfnamefont {G.~J.}\ \bibnamefont
  {Erickson}},\ and\ \bibinfo {editor} {\bibfnamefont {P.~O.}\ \bibnamefont
  {Neudorfer}}}\ (\bibinfo  {publisher} {Springer},\ \bibinfo {address}
  {Dordrecht, Netherlands},\ \bibinfo {year} {1992})\ pp.\ \bibinfo {pages}
  {1--21}\BibitemShut {NoStop}%
\bibitem [{\citenamefont {Garcia-Millan}\ \emph {et~al.}(2024)\citenamefont
  {Garcia-Millan}, \citenamefont {Schüttler}, \citenamefont {Cates},\ and\
  \citenamefont {Loos}}]{garciamillan2024optimalclosedloopcontrolactive}%
  \BibitemOpen
  \bibfield  {author} {\bibinfo {author} {\bibfnamefont {R.}~\bibnamefont
  {Garcia-Millan}}, \bibinfo {author} {\bibfnamefont {J.}~\bibnamefont
  {Schüttler}}, \bibinfo {author} {\bibfnamefont {M.~E.}\ \bibnamefont
  {Cates}},\ and\ \bibinfo {author} {\bibfnamefont {S.~A.~M.}\ \bibnamefont
  {Loos}},\ }\bibfield  {title} {\bibinfo {title} {Optimal closed-loop control
  of active particles and a minimal information engine},\ }\Eprint
  {https://arxiv.org/abs/arXiv:2407.18542} {arXiv:2407.18542}  (\bibinfo {year}
  {2024}),\ \bibinfo {note} {preprint}\BibitemShut {NoStop}%
\bibitem [{\citenamefont {Sezik}\ \emph {et~al.}(2025)\citenamefont {Sezik},
  \citenamefont {Knight}, \citenamefont {Alston}, \citenamefont {Roberts},
  \citenamefont {Bertrand}, \citenamefont {Pruessner},\ and\ \citenamefont
  {Cocconi}}]{SezikKnightAlstonCocconi:2025b}%
  \BibitemOpen
  \bibfield  {author} {\bibinfo {author} {\bibfnamefont {E.}~\bibnamefont
  {Sezik}}, \bibinfo {author} {\bibfnamefont {J.}~\bibnamefont {Knight}},
  \bibinfo {author} {\bibfnamefont {H.}~\bibnamefont {Alston}}, \bibinfo
  {author} {\bibfnamefont {C.}~\bibnamefont {Roberts}}, \bibinfo {author}
  {\bibfnamefont {T.}~\bibnamefont {Bertrand}}, \bibinfo {author}
  {\bibfnamefont {G.}~\bibnamefont {Pruessner}},\ and\ \bibinfo {author}
  {\bibfnamefont {L.}~\bibnamefont {Cocconi}},\ }\bibfield  {title} {\bibinfo
  {title} {Hidden state inference in drift-diffusive processes via conditional
  splitting probabilities}} (\bibinfo {year} {2025}),\ \bibinfo {note} {in
  preparation}\BibitemShut {NoStop}%
\bibitem [{\citenamefont {Kardar}(2000)}]{Kardar:2000}%
  \BibitemOpen
  \bibfield  {author} {\bibinfo {author} {\bibfnamefont {M.}~\bibnamefont
  {Kardar}},\ }\bibfield  {title} {\bibinfo {title} {Stochastic dynamics of
  growing films},\ }in\ \href
  {https://www.worldscientific.com/doi/abs/10.1142/4613} {\emph {\bibinfo
  {booktitle} {Annual Reviews of Computational Physics}}},\ Vol.\ \bibinfo
  {volume} {VIII},\ \bibinfo {editor} {edited by\ \bibinfo {editor}
  {\bibfnamefont {D.}~\bibnamefont {Stauffer}}}\ (\bibinfo  {publisher} {World
  Scientific},\ \bibinfo {address} {Singapore},\ \bibinfo {year} {2000})\ pp.\
  \bibinfo {pages} {1--47}\BibitemShut {NoStop}%
\bibitem [{\citenamefont {{Le Bellac}}(1991)}]{LeBellac:1991}%
  \BibitemOpen
  \bibfield  {author} {\bibinfo {author} {\bibfnamefont {M.}~\bibnamefont {{Le
  Bellac}}},\ }\href@noop {} {\emph {\bibinfo {title} {Quantum and Statistical
  Field Theory [Phenomenes critiques aux champs de jauge, English]}}}\
  (\bibinfo  {publisher} {Oxford University Press},\ \bibinfo {address} {New
  York, NY, USA},\ \bibinfo {year} {1991})\ \bibinfo {note} {translated by G.
  Barton}\BibitemShut {NoStop}%
\bibitem [{\citenamefont {{van Kampen}}(1992)}]{vanKampen:1992}%
  \BibitemOpen
  \bibfield  {author} {\bibinfo {author} {\bibfnamefont {N.~G.}\ \bibnamefont
  {{van Kampen}}},\ }\href@noop {} {\emph {\bibinfo {title} {Stochastic
  Processes in Physics and Chemistry}}}\ (\bibinfo  {publisher} {Elsevier
  Science B. V.},\ \bibinfo {address} {Amsterdam, The Netherlands},\ \bibinfo
  {year} {1992})\ \bibinfo {note} {third impression 2001, enlarged and
  revised}\BibitemShut {NoStop}%
\bibitem [{\citenamefont {Doi}(1976)}]{Doi:1976}%
  \BibitemOpen
  \bibfield  {author} {\bibinfo {author} {\bibfnamefont {M.}~\bibnamefont
  {Doi}},\ }\bibfield  {title} {\bibinfo {title} {Second quantization
  representation for classical many-particle system},\ }\href
  {http://stacks.iop.org/0305-4470/9/1465} {\bibfield  {journal} {\bibinfo
  {journal} {J. Phys. A: Math. Gen.}\ }\textbf {\bibinfo {volume} {9}},\
  \bibinfo {pages} {1465} (\bibinfo {year} {1976})}\BibitemShut {NoStop}%
\bibitem [{\citenamefont {Peliti}(1985)}]{Peliti:1985}%
  \BibitemOpen
  \bibfield  {author} {\bibinfo {author} {\bibfnamefont {L.}~\bibnamefont
  {Peliti}},\ }\bibfield  {title} {\bibinfo {title} {Path integral approach to
  birth-death processes on a lattice},\ }\href@noop {} {\bibfield  {journal}
  {\bibinfo  {journal} {J. Phys. (Paris)}\ }\textbf {\bibinfo {volume} {46}},\
  \bibinfo {pages} {1469} (\bibinfo {year} {1985})}\BibitemShut {NoStop}%
\end{thebibliography}%

\clearpage
\newpage
\mbox{~}

\onecolumngrid

\begin{center}
{\large\bf Supplementary Material}
\end{center}



\renewcommand{\theequation}{S\arabic{equation}}

\renewcommand{\thefigure}{S\arabic{figure}}

\renewcommand{\thesection}{S\Roman{section}}

\tableofcontents

\section{Notation and basic formulae}\seclabel{SM:notation}
This section provides technical details to accompany the derivations in the main text and \Sref{calc_EPR}. We consider three different ensembles of paths with four different probability weights. Firstly, the thermal noise $\xi(t)$ for $t\in[0,T]$ is Gaussian white noise, independent of the self-propulsion $w(t)$ and with vanishing mean. Its path density is \cite{Kardar:2000}
\begin{equation}\elabel{xipath_prob}
 \PprobXI[\xipath] \propto  
 \Exp{-\frac{1}{4D}\int_0^T\dint{t}\xi(t)^2}
\end{equation}
up to a normalisation that need not be further specified. The expectation under this path density is written 
\begin{equation}
    \XIave{\bullet} = \int \Dint{\xi} \bullet \PprobXI[\xipath]
\end{equation}
so that normalisation produces $\XIave{1}=1$. The Gaussian noise is fully described by $\XIave{\xi(t)}=0$ and $\XIave{\xi(t)\xi(t')}=2D\delta(t-t')$ for any $t,t'\in[0,T]$.

Similarly, the self-propulsion ``noise'' is the dimensionless $w(t)$, which has path density $\PprobW[\wpath]$ and expectation $\Wave{\bullet}$, so that, again $\Wave{1}=1$ and $\nu\Wave{w(t)}$ is the expected self-propulsion velocity of the active particle, \Eref{langevin}. 

We further require a modified path density, \Erefs{overline_def_main} and \eref{overline_def}, 
\begin{equation}\elabel{def_PprobWSTAR}
    \PprobWSTAR[\wpath] = \frac{1}{\NormWSTAR(\nu)} \Exp{-\frac{\nu^2}{4D} \int_0^T\dint{t} w(t)^2 }
    \;\PprobW[\wpath]
\end{equation}
where we made the normalisation $\NormWSTAR(\nu)$ explicit, because it will need to cancel in \Eref{partial_EP_2}. The expectation under $\PprobWSTAR[\wpath]$ is $\WSTARave{\bullet}$, \Eref{overline_def}, so that $\WSTARave{1}=1$. 
To leading order $\nu^0$ in the exponential, $\WSTARave{\bullet}$ coincides with $\Wave{\bullet}$, \Eref{def_bave}.

We finally introduce the path density $\PprobX[\xpath]$, whose expectations are written as $\Xave{\bullet}$, \Eref{simplified_Xave}. For the most part, we may think of $\xpath$ as a functional of $\wpath$ and $\xipath$, but because of the simple, additive form of the Langevin \Eref{langevin}, any two of $\xpath, \wpath, \xpath$ are ``linearly independent'' and suffice to construct the third. We will use, in particular, \Eref{simplified_Xave},
\begin{equation}\elabel{Pprob_xi_w_factorising}
\PprobXW[\xpath,\wpath]
\ \Dint{x}
\ \Dint{w}
=\PprobXI[\xipath,\wpath]
\ \Dint{\xi}
\ \Dint{w}
=\PprobXI[\xipath]\ \Dint{\xi}
\PprobW[\wpath]\ \Dint{w}
\end{equation}
where the last equality is due to the statistical independence of $\xipath$ and $\wpath$.

We finally list the three different entropy productions we consider. All are based on the Kullback-Leibler divergence of the path densities \cite{gaspard2004time,cocconi2020entropy}. For the joint path $\xpath,\wpath$ this is \Eref{EP_full}
\begin{equation}\elabel{EP_full_SM}
    \EPRfull = \lim_{T \to \infty} \frac{1}{T} \int\Dint{x}\!\!\int\Dint{w}\,\PprobXW[\xpath,\wpath]
    \ln\left(\frac{\PprobX[\xpath, \wpath]}{\PprobXW[\Rxpath, \Rwpath]}\right)\;,
\end{equation}
which factorises in time, as $x(t),w(t)$ is Markovian. As a result, there is no need to consider large $T$ and the (internal) entropy production formally becomes
\begin{equation}\elabel{EP_full_SM_factorised}
    \EPRfull = \int 
    \dint{x_0}\dint{x_1}
    \dint{w_0}\dint{w_1}
    P(x_0,w_0) \ \mathring{W}(x_0,w_0\to x_1,w_1)
    \ln \left(
\frac{\mathring{W}(x_0,w_0\to x_1,w_1)}{\mathring{W}(x_1,w_1\to x_0,w_0)}
    \right)
\end{equation}
where $P(x_0,w_0)$ denotes the probability density for the system to be found at $x_0,w_0$ in the steady state and $\mathring{W}(x_0,w_0\to x_1,w_1)$ is the Poissonian transition rate from $x_0,w_0$ to $x_1,w_1$, representing operators as $x,w$ are continuous variables \cite{PruessnerGarcia-Millan:2022}.

The entropy production of $\xpath$ alone, however, does not simplify in this way, \Eref{partial_EP_defn},
\begin{equation}\elabel{partial_EP_defn_SM_again}
    \EPRpartX = \lim_{T \to \infty} \frac{1}{T} \int \Dint{x}\;\PprobX[\xpath]\ln\left(\frac{\PprobX[\xpath]}{\PprobX[\Rxpath]}\right)\;.
\end{equation}
This is the quantity we focus on in the present work. Similarly,
\begin{equation}\elabel{partial_EPW_defn_SM}
    \EPRpartW = \lim_{T \to \infty} \frac{1}{T} \int \Dint{w}\;\PprobW[\wpath]\ln\left(\frac{\PprobW[\wpath]}{\PprobW[\Rwpath]}\right)\;,
\end{equation}
is the entropy production rate (EPR) of the $w$-process on its own, as used in \Tref{processes}. Even when the path-probabilities featuring in \Eref{partial_EPW_defn_SM} may in principle be difficult to obtain, if $w$ is itself Markovian, \Eref{partial_EPW_defn_SM} factorises just like \Eref{EP_full_SM} does to \Eref{EP_full_SM_factorised}.

\section{Detailed derivation of the Partial Entropy Production Rate} \seclabel{calc_EPR}
In this section we derive the central Eq.~\eqref{eq:partial_EP_expansion_free} using the Onsager-Machslup path integral formalism and provide a detailed derivation of the leading order \Erefs{EPR_leading_order_main} and \eref{EPR_leading_order_SM} together with the relevant symmetry arguments. We work in the symmetric Stratonovich interpretation whenever such a choice has to be made. The starting point is the partial entropy production rate (EPR) in \Erefs{partial_EP_defn} and \Eref{partial_EP_defn_SM_again}
\begin{equation}\elabel{partial_EP_defn_SM}
    \EPRpartX = \lim_{T \to \infty} \frac{1}{T} \int \Dint{x}\PprobX[\xpath]\ln\left(\frac{\PprobX[\xpath]}{\PprobX[\Rxpath]}\right)\;.
\end{equation}
The path probability $\mathcal{P}[\{x(t)\}]$ can be expressed as the full path probability $\mathcal{P}[\{x(t)\}, \{w(t)\}]$ marginalised over $w(t)$, \Eref{Pprobx_from_conditional}
\begin{equation}\elabel{Pprob_xw_marginalised}
\begin{split}
    \PprobX[\xpath]&=\int\Dint{w}
    \PprobXW[\xpath,\wpath]\\
    &=\int\Dint{w}
    \PprobX[\xpath|\wpath] \PprobW[\wpath]\;,
    \end{split}
\end{equation}
which we may use to rewrite \Eref{partial_EP_defn_SM} as 
\begin{equation}\elabel{partial_EP_1}
    \EPRpartX = \lim_{T \to \infty} \frac{1}{T} \int \Dint{x}
    \int\Dint{w'}\PprobX[\xpath|\wdpath]\;\PprobW[\wdpath]\;
    \ln\left(\frac{
    \int\Dint{w''}\PprobX[\xpath|\wdddpath]\;\PprobW[\wddpath]
    }{
    \int\Dint{w'''}\PprobXW[\Rxpath|\wdddpath]\;\PprobW[\wdddpath]
    }
    \right)
\end{equation}
where the primes in $w', w'', w'''$ are introduced to emphasise that these are independent dummy variables. 
This is of particular importance when marginalising the path probability $\PprobXW[\Rxpath,\Rwpath]$ in the denominator, which may more naturally be written as $\int\Dint{w^R}\PprobXW[\Rxpath|\Rwpath]\PprobW[\Rwpath]$, where, however, $w^R(t)=w(T-t)$ is in fact only a dummy variable, say $w'''(t)$.

The conditional path probability of trajectories $\PprobX[\xpath|\wpath]$ in \Eref{Pprob_xw_marginalised} can be expressed using the Onsager-Machlup formalism \cite{Taeuber:2014},
\begin{equation}\elabel{fulldist}
    \mathcal{P}\left[\{x(t)\}|\{w(t)\}\right] =\frac{1}{\NC} \exP \left\{-\frac{1}{4D} \int_0^T \dint{t} \left(\dot{x}(t) -\nu w(t)\right)^2
    \right\}\;,
\end{equation}
using the Langevin \Eref{langevin} and the statistical weight of the noise path $\xi(t)$, \Eref{xipath_prob}. To avoid ambiguity, we state in \Eref{fulldist} explicitly a normalisation constant $1/\NC$. Using \Eref{Pprob_xw_marginalised} now yields the path probability of $\{x(t)\}$:
\begin{equation}\elabel{path_prob_x}
     \PprobX[\xpath]=\frac{1}{\NC}
     \Exp{-\frac{1}{4D}\int_0^T \dint{t'}
     \left( \xdot(t') \right)^2 }
\int\Dint{w} 
\Exp{-\frac{1}{4D} 
\int_0^T \dint{t'}
\left[-2\nu w(t') \xdot(t') + (\nu w(t'))^2 \right]
}
\PprobW[\wpath] 
\end{equation}

The path probability $\PprobX[\Rxpath]$ is the probability of observing the reverse of $\xpath$. It is obtained from \Eref{path_prob_x} with the slight subtlety that the ``velocity'' $\xdot(t')$ refers to the derivative $\plaind x(t)/\plaind t$ evaluated at $t=t'$, the dummy variable in the integral of \Eref{path_prob_x}. In the reverse path, the velocity is indeed reversed,
\begin{equation}
\left.\ddt\right|_{t=t'} x(T-t) = -\xdot(T-t')
\end{equation}
so that 
\begin{align}\elabel{path_prob_reverse_x}
     &\PprobX[\Rxpath]\\
     &=\frac{1}{\NC} 
     \Exp{-\frac{1}{4D}\int_0^T \dint{t'}
     \left( - \xdot(T-t') \right)^2 }
\int \Dint{w} 
\Exp{-\frac{1}{4D} 
\int_0^T \dint{t'}
\left[2\nu w(t') \xdot(T-t') + (\nu w(t'))^2 \right]
}
\PprobW[\wpath] \nonumber \\
&=\frac{1}{\NC}\Exp{-\frac{1}{4D}\int_0^T \dint{t''}
     \left( \xdot(t'') \right)^2 }
\int \Dint{w} 
\Exp{-\frac{1}{4D} 
\int_0^T \dint{t''}
\left[2\nu w(T-t'') \xdot(t'') + (\nu w(T-t''))^2 \right]
}
\PprobW[\wpath]\nonumber 
\end{align}
using $t''=T-t'$ as the dummy variable.

We proceed to calculate the ratio inside the logarithm of \Eref{partial_EP_defn_SM} using \Erefs{path_prob_x} and \Eref{path_prob_reverse_x}. The $w(t)$-independent pre-factors cancel, yielding
\begin{equation}\elabel{partial_EP_2} 
    \begin{split}
    \frac{\PprobX[\xpath]}{\PprobX[\Rxpath]}
    &=
\frac
{
\int \Dint{w} 
\Exp{-\frac{1}{4D} 
\int_0^T \dint{t'}
\left[-2\nu w(t') \xdot(t') + (\nu w(t'))^2 \right]
}
\PprobW[\wpath]
}
{
\int \Dint{w} 
\Exp{-\frac{1}{4D} 
\int_0^T \dint{t''}
\left[2\nu w(T-t'') \xdot(t'') + (\nu w(T-t''))^2 \right]
}
\PprobW[\wpath]
}\\
&=
\frac
{\NormWSTAR(\nu)
\WSTARave{
\Exp{\frac{\nu}{2D} 
\int_0^T \dint{t}
w(t) \xdot(t)
}}}
{
\phantom{-}\NormWSTAR(\nu)\WSTARave{
\Exp{-\frac{\nu}{2D} 
\int_0^T \dint{t}
w(T-t) \xdot(t)
}}\phantom{-}}
\; ,
    \end{split}
\end{equation}
where $\WSTARave{\bullet}$ represents an average taken with respect to the starred, steady-state $w$-path probability distribution $\PprobWSTAR$ of \Eref{def_PprobWSTAR},
\begin{equation}\elabel{overline_def} 
    \WSTARave{\bullet}
    =\frac{1}{\NormWSTAR(\nu)}\int\Dint{w}\bullet 
    \Exp{-\frac{\nu^2}{4D} \int_0^T\dint{t} w(t)^2 }
    \;\PprobW[\wpath] 
    =
    \int\Dint{w}\bullet \PprobWSTAR[\wpath] \;,
\end{equation}
to account for the additional statistical weight $\exp{-\nu^2 \int_0^T\dint{t} w(t)^2/(4D)}$. Crucially, the normalisation $\NormWSTAR(\nu)$ in the numerator and the denominator of \Eref{partial_EP_2} in the second line, which is not present in the first line, does not depend on the direction of time in which $w(t)^2$ is integrated, so that it cancels. Further, if $w(t)^2$ is constant, as is the case for telegraphic noise alternating between $w_+$ and $w_-=-w_+$, then expectations under $\PprobWSTAR[\wpath]$ are identical to those under $\PprobW[\wpath]$ and many of the arguments below, concerning the leading order in $\nu$, simplify dramatically.

The definition \Eref{def_PprobWSTAR} of $\PprobWSTAR[\wpath]$ ensures that moments and cumulants taken with respect to $\PprobWSTAR[\wpath]$ will be equal to those taken with respect to $\PprobW[\wpath]$ to leading, \ie zeroth, order in $\nu^2/D$ and $\WSTARave{\bullet}=\Wave{\bullet}+\OC(\nu^2/D)$. Further, $\PprobWSTAR[\wpath]$ and $\PprobW[\wpath]$ share the same symmetries, because the additional statistical weight $\exp{-\nu^2\int\dint{t}w^2/(4D)}$ is $\SymParityTime$-symmetric.

The logarithm of \Eref{partial_EP_2} can be recognised as the difference between two cumulant generating functions \cite{LeBellac:1991} of the form 
\begin{subequations}
\elabel{logs_as_generatingfunctions}
\begin{align}
    \ln \left(\ \WSTARave{
    \Exp{\frac{\nu}{2D} 
    \int\dint{t} w(t) \xdot(t)} }\ \right)
    &= \sum_{n=1}^\infty \frac{1}{n!}\left(\frac{\nu}{2D}\right)^n \int_0^T 
    \dint{t_1}\!\!\sdots\dint{t_n}
    \; \xdot(t_1)\ldots\xdot(t_n)
\  \WSTARaveS{w(t_1)\ldots w(t_n)}{\cbb}\;,\\ 
    \ln \left(\ \WSTARave{
    \Exp{-\frac{\nu}{2D} 
    \int\dint{t} w(T-t) \xdot(t)} }\ \right)
    &= \sum_{n=1}^\infty \frac{1}{n!}\left(\frac{-\nu}{2D}\right)^n \int_0^T 
    \dint{t_1}\!\!\sdots\dint{t_n}
    \; \xdot(t_1)\ldots\xdot(t_n)
\  \WSTARaveS{w(T-t_1)\ldots w(T-t_n)}{\cbb}\;,
\end{align}
\end{subequations}
where the superscript $\cbb$ of $\WSTARaveS{\bullet}{\cbb}$ denotes the cumulant of $\bullet$ with respect to $\PprobWSTAR[\wpath]$. Using \Eref{logs_as_generatingfunctions} for the logarithm of \Eref{partial_EP_2} in \Eref{partial_EP_defn_SM}, we arrive at 
\begin{equation}\elabel{partial_EP_expansion_1}
\begin{split}
    \EPRpartX &= \lim_{T \to \infty} \frac{1}{T} \int \Dint{x}\PprobX[\xpath]
    \sum_{n=1}^\infty \frac{1}{n!}\left(\frac{\nu}{2D}\right)^n \int_0^T 
    \dint{t_1}\!\!\sdots\dint{t_n}
    \; \xdot(t_1)\ldots\xdot(t_n)\\
& \pushright{\times
\Big( \WSTARaveS{w(t_1)\ldots w(t_n)}{\cbb}
-
(-)^n \WSTARaveS{w(T-t_1)\ldots w(T-t_n)}{\cbb}
\Big)} \ ,
    \end{split}
\end{equation}
which can be further symmetrised by changing the dummy variables $t_1,\ldots,t_n$ to $t_1'=T-t_1,\ldots,t_n'=T-t_n$ and then relabelling, so that
\begin{equation}\elabel{partial_EP_expansion_symmetric}
\begin{split}
    \EPRpartX 
    &= \lim_{T \to \infty} \frac{1}{2T} 
    \sum_{n=1}^\infty \frac{1}{n!}\left(\frac{\nu}{2D}\right)^n 
    \int_0^T 
    \dint{t_1}\!\!\sdots\dint{t_n}
    \; 
    \Big( \Xave{\xdot(t_1)\ldots \xdot(t_n)} -
    (-)^n \Xave{\xdot(T-t_1)\ldots \xdot(T-t_n) } \Big)\\
&\pushright{\times \Big( \WSTARaveS{w(t_1)\ldots w(t_n)}{\cbb}
-
(-)^n \WSTARaveS{w(T-t_1)\ldots w(T-t_n)}{\cbb}
\Big)} \ .
    \end{split}
\end{equation}
where we have used the shorthand
\begin{equation}\elabel{simplified_Xave}
    \Xave{\bullet} = \int\Dint{x} \bullet \PprobX[\xpath] 
    = \int\Dint{x}\;\Dint{w} \bullet \PprobXW[\xpath,\wpath] 
    = \int\Dint{\xi}\;\Dint{w} \bullet \PprobXI[\xipath] \PprobW[\wpath] 
    = \XIave{\Wave{\bullet}}
    = \Wave{\XIave{\bullet}}
\end{equation}
for expectations under $\PprobX[\xpath]$ as $\PprobXW[\xpath,\wpath]\ \Dint{x}\ \Dint{w}=\PprobXI[\xipath,\wpath]\ \Dint{\xi}\ \Dint{w}$, \Eref{Pprob_xi_w_factorising}, and the latter probability factorises since $\xi(t)$ and $w(t)$ are independent. 

\Erefs{partial_EP_expansion_symmetric} and \eref{partial_EP_expansion_free} are the key result of the present work. What follows are general arguments about the expectations $\Xave{\xdot\ldots}$ and how symmetries determine the lowest order $\nlowest$ with which the $\WSTARaveS{w\ldots}{\cbb}$ enter.

\subsection{Velocity correlators \texorpdfstring{$\Xave{\xdot(t_1)\ldots}$}{}}
Using the Langevin \Eref{langevin}, the moments $\Xave{\xdot(t_1)\ldots \xdot(t_n)}$ can immediately be expressed in terms of moments of $w(t)$ as well as the noise correlator, for example
\begin{multline}\elabel{xdot3_example}
\Xave{\xdot(t_1)\xdot(t_2)\xdot(t_3)} \\
= 
\int\Dint{w} \Big(
\nu^3 w(t_1)w(t_2)w(t_3) 
+ 2D\nu\delta(t_1-t_2)w(t_3) 
+ 2D\nu\delta(t_2-t_3)w(t_1) 
+ 2D\nu\delta(t_3-t_1)w(t_2)
\Big)\PprobW[\wpath]\\
=
\nu^3\Wave{w(t_1)w(t_2)w(t_3)}
+2D\nu \Big(\delta(t_1-t_2)+\delta(t_2-t_3)+\delta(t_3-t_1)\Big) \Wave{w}
\end{multline}
using the correlator \Eref{noise_correlator} of the thermal noise $\xi(t)$ in the first equality and the notation $\Wave{\bullet}$ to take an expectation with respect to $\PprobW[\wpath]$ rather than $\PprobWSTAR[\wpath]$, \Eref{overline_def},
\begin{equation}\elabel{def_bave}
\Wave{\bullet}=\int\Dint{w}\bullet\PprobW[\wpath] \;.
\end{equation}
In the steady state $\Wave{w(t_1)}=\Wave{w(t_2)}=\Wave{w(t_3)}=\Wave{w}$.

It is obvious that the moments $\Xave{\xdot(t_1)\ldots \xdot(t_n)}$ are identical to $\nu^n\Wave{w(t_1)\ldots w(t_n)}$ to order $\nu^n$, but that they also contain lower order contributions, with $\nu^2 w(t_i)w(t_j)$ effectively replaced by $2D\delta(t_i-t_j)$, \Eref{noise_correlator}. However, in the following we will demonstrate that \emph{if the lowest contributing order due to the cumulant term in \Eref{partial_EP_expansion_symmetric} is $\nlowest$ then the $\nlowest$-time correlators $\Xave{\xdot(t_1)\sdots\xdot(t_\nlowest)}$ contribute exclusively to order $\nu^\nlowest$}, and both cumulants and correlators amount to essentially the same term, resulting in \Eref{EPR_leading_order_main} or \Eref{EPR_leading_order_SM} below.

To arrive at this result we will consider the cumulants $\WSTARaveS{w(t_1)\ldots w(t_n)}{\cbb}$ and show how their properties affect the moments $\Wave{w(t_1)\ldots w(t_n)}$ and therefore those of $\Xave{\xdot(t_1)\ldots\xdot(t_n)}$. To this end, we introduce the shorthands
\begin{subequations}\elabel{def_corrStarETAL}
\begin{align}
   \corrStar_n&= \WSTARaveS{w(t_1)\ldots w(t_n)}{\cbb} - (-)^n \WSTARaveS{w(T-t_1)\ldots w(T-t_n)}{\cbb}\\
   W_n&=\Wave{w(t_1)\ldots w(t_n)} - (-)^n \Wave{w(T-t_1)\ldots w(T-t_n)}\\
   X_n&=\Xave{\xdot(t_1)\ldots\xdot(t_n)} - (-)^n \Xave{\xdot(T-t_1)\ldots\xdot(T-t_n)}
\;.
\end{align}
\end{subequations}
We assume that $\corrStar_n$ has been found to vanish for all $n<\nlowest$. Crucially, $\corrStar_n$ at $\nu=0$, which we denote by $\corrStar[\nu=0]_n$, is assumed to vanish for all $n<\nlowest$. In fact, in \Sref{symmetries} we will find precisely such $\nlowest$ so that $\corrStar_n$ vanishes for all $\nu$ and $n<\nlowest$.

Firstly, $\corrStar_n=0$ for $\nu=0$ and $n<\nlowest$ implies that $W_n$ vanishes for all $n<\nlowest$, as moments can be written in terms of sums of products of cumulants. Explicit expressions are immediately found as the moment generating function is the exponential of the cumulant generating function. Were it not for the additional $(-)^n$ in \Eref{def_corrStarETAL}, this would be a matter of cumulants and thus moments being invariant under time reversal. To make the argument in the presence of $(-)^n$, we write $\corrStar[\nu=0]_n=0$ for $n<\nlowest$ as
\begin{equation}
    \WaveS{w(t_1)\ldots w(t_n)}{\cbb} = (-)^n \WaveS{w(T-t_1)\ldots w(T-t_n)}{\cbb} \;.
\end{equation}
The sign $(-)^n$ on the right is easily accounted for by noticing that any product of cumulants contributing to $W_n$ with odd $n$ contains an odd number of odd cumulants and any contribution to even $n$ contains an even number of odd moments. Any moment $\Wave{w(t_1)\sdots w(t_n)}$ can thus be shown to equal $\Wave{w(T-t_1)\sdots w(T-t_n)}$ up to a sign $(-)^n$. In other words, $W_n=0$ for $n<\nlowest$.

Secondly, if the moments $W_n$ vanish for $n<\nlowest$, then the $X_n$ vanish as well. The argument works along the same lines as above, with the added difficulty of the $\delta$-functions due to the thermal noise. Those are, however, even functions of their argument and invariant under time reversal, so that they cancel suitably across forward and time-reversed terms, \eg \Eref{xdot3_example}. Any $X_n$ can thus be written as a sum of products of $\delta$-functions and $W_\ell$ with $\ell<n$, which all vanish by assumption. In other words, $X_n=0$ for $n<\nlowest$.

Thirdly, if the moments $W_n$ vanish for $n<\nlowest$, then $X_m$ for $m\ge\nlowest$ are at least of order $\nu^\nlowest$. This can be seen by noticing that any term of order $\nu^n$ with $n<\nlowest$ appearing in $X_m$ can be written as a product of $\delta$-functions multiplying $W_n$, which, however, vanishes. The only term of order $\nlowest$ in $X_\nlowest$ is $\nu^\nlowest W_\nlowest$, \ie $X_\nlowest=\nu^\nlowest W_\nlowest$ and $X_m\in\OC(\nu^\nlowest)$ for $m>\nlowest$.

We finally observe that $\corrStar_\nlowest$ is to lowest order $\corrStar[\nu=0]_\nlowest$ with corrections at least of order $\nu^2/D$, \Eref{overline_def}. Writing all cumulants appearing in $\corrStar[\nu=0]_\nlowest$ in terms of moments $\Wave{w(t_1)\ldots w(t_n)}$ and thus in terms of $W_n$ leaves only one non-vanishing term, namely $W_\nlowest$, \ie  $\corrStar_\nlowest=W_\nlowest + \OC(\nu^2/D)$

Returning to \Eref{partial_EP_expansion_symmetric} and writing it in the form 
\begin{equation}\elabel{EPR_in_X_and_corrStar}
    \EPRpartX 
    = \lim_{T \to \infty} \frac{1}{2T} 
    \sum_{n=1}^\infty \frac{1}{n!}\left(\frac{\nu}{2D}\right)^n 
    \int_0^T 
    \dint{t_1}\!\!\sdots\dint{t_n}
    \; X_n \corrStar_n
\ ,
\end{equation}
the lowest order term in $\nu$ is thus \Eref{EPR_leading_order_main}
\begin{equation}
\elabel{EPR_leading_order_SM}
    \EPRpartX 
    = \lim_{T \to \infty} \frac{1}{2T} 
    \frac{1}{\nlowest!}\left(\frac{\nu^2}{2D}\right)^\nlowest 
    \int_0^T 
    \dint{t_1}\!\!\sdots\dint{t_n}
    \; \Big(
    \Wave{w(t_1)\ldots w(t_\nlowest)}
    - (-)^\nlowest
    \Wave{w(T-t_1)\ldots w(T-t_\nlowest)}
    \Big)^2
 + \OC\left(\frac{\nu^{2(\nlowest+1)}}{D^{\nlowest+1}}\right) \ ,
\end{equation}
where both, cumulants $\WSTARaveS{w(t_1)\ldots w(t_n)}{\cbb}$ and moments $\Xave{\xdot(t_1)\ldots\xdot(t_n)}$, are to leading order captured by the same term, $\nu^\nlowest W_\nlowest$.

To determine the next order correction as stated in \Eref{EPR_leading_order_SM}, we need to consider corrections to $X_\nlowest$ and $\corrStar_\nlowest$, as well as $X_{\nlowest+1}$ and $\corrStar_{\nlowest+1}$. As for $X_\nlowest$, there are no corrections to $\nu^\nlowest W_\nlowest$, while $\corrStar_\nlowest$ is corrected by  powers of $\nu^2/D$, giving rise to a term $\OC(\nu^{2\nlowest+2}/D^{\nlowest+1})$ overall. At the next order in the summation, $\nlowest+1$, the correlator $X_{\nlowest+1}$ is $\OC(\nu^{\nlowest+1})$ as there cannot be a contribution from $\XIave{\xi\xi}$, \Eref{noise_correlator}, yet, while $\corrStar_{\nlowest+1}$ enters with a pre-factor $(\nu/D)^{\nlowest+1}$ into $\EPRpartX$, so that the overall correction is $\propto(\nu^2/D)^{\nlowest+1}$, the same order as $\OC(\nu^{2\nlowest+2}/D^{\nlowest+1})$. In general, we thus state the next order as $\OC(\nu^{2\nlowest+2}/D^{\nlowest+1})$.

In the examples in \Sref{leading_order}, the next orders to consider in the summation are $\nlowest+2$, where $X_{\nlowest+2}$ consists of terms $\OC(\nu^{\nlowest+2})$ and $\OC(D\nu^{\nlowest})$, due to $\XIave{\xi\xi}$, \Eref{noise_correlator}, and $\corrStar_{\nlowest+2}$ enters with pre-factor $(\nu/D)^{\nlowest+2}$, producing corrections of order
$\nu^{2\nlowest+4}/D^{\nlowest+2}$
and
$\nu^{2\nlowest+2}/D^{\nlowest+1}$ overall to next order in the summation. The latter is of lower order and the same as the one determined above and thus the one we state.

Correlation functions $\Wave{w(t_1)\sdots w(t_n)}$ are often stated with a given time-ordering, as any time order can be achieved by permutation of arguments. If the integrand is free of $\delta$-functions, the integral in \Eref{EPR_leading_order_SM} is most conveniently carried out using the fixed, given time ordering, say $t_n>t_{n-1}>\ldots>t_1$. One can easily show that
\begin{equation}\elabel{simplified_integral}
    \int_0^T \dint{t_1}\!\!\sdots\dint{t_n} \ f(t_1,\sdots,t_n)
    =
    (n!) 
    \int_0^T \dint{t_1}
    \int_{t_1}^T \dint{t_2}
    \int_{t_2}^T \dint{t_3}
\ldots
    \int_{t_{n-1}}^T \dint{t_n}
    f(t_1,\sdots,t_n) 
\end{equation}
for any integrand $f(t_1,\sdots,t_n)$ that is invariant under permutations of the argument
for example by using a sum of all indicator functions of all time-orderings and then relabelling the dummy-variables to reduce the integral to one that maintains $t_n>t_{n-1}>\ldots>t_1$, over the sum $f(t_1,t_2,\sdots,t_n)+f(t_2,t_1,\sdots,t_n)+\ldots=(n!)f(t_1,\sdots,t_n)$.

\subsection{Symmetries determining the lowest relevant order \texorpdfstring{$\nlowest$}{n*}}\seclabel{symmetries}
We briefly summarise the arguments about the lowest non-vanishing order $\nlowest$, of $\corrStar_n$ and $W_n$, \Eref{def_corrStarETAL}.

If the first order $W_1=2\Wave{w}$ does not vanish, then the self-propulsion results in net-transport. Time-reversal symmetry is ``trivially'' and very apparently broken, $\EPRpartX=(\nu\Wave{w})^2/D + \OC(\nu^3/D^2)$.

The non-trivial case, $\Wave{w}=0$ and therefore $W_1=0$, is somewhat subtle, as it does not imply $\corrStar_1=0$, but rather $\corrStar_1\in\OC(\nu^2/D)$. However, the premise in the preceding section was that $\corrStar_n$ vanishes for all $\nu$ and $n<\nlowest$. And yet, the arguments presented to derive \Eref{EPR_leading_order_SM} remain valid, as the $n=1$-term in \Eref{EPR_in_X_and_corrStar} is suppressed by $W_1=0$ and $\corrStar[\nu=0]_1=0$ indeed holds, which is sufficient, as suggested after \Eref{def_corrStarETAL}. As such, $\Wave{w}=0$ implies that the lowest order contribution cannot be at first order, \textit{i.e.} $\nlowest\ne1$.

The second order, $\corrStar_2$, cannot possibly contribute in the steady state, as $\WSTARave{w(t_1)w(t_2)}$ depends only on $|t_1-t_2|$ and thus is $\SymTime$-symmetric, so that $\corrStar_2=0$. It follows that the lowest relevant order cannot be $2$, \ie $\nlowest\ne2$ generally.

Beyond second order, $\nlowest$ is determined by the symmetries of $\corrStar_n$. It vanishes for odd $n$ whenever $\PprobW[\wpath]$ and therefore $\PprobWSTAR[\wpath]$ is $\SymParity$-symmetric, \ie $\PprobWSTAR$ is an even function in $w$, so that $\PprobWSTAR[\wpath]=\PprobWSTAR[\Mwpath]$. As a result all odd moments vanish, as $\SymParity$-symmetric $\PprobW[\wpath]$ implies for any $n$
\begin{equation}\elabel{even_Pprob}
\begin{split}
\WSTARave{w(t_1)\ldots w(t_n)}&=
\int\Dint{w} w(t_1)\sdots w(t_n) \PprobWSTAR[\wpath] = (-)^n \int\Dint{w'} w'(t_1)\sdots w'(t_n) \PprobWSTAR[\Mwdpath]\\
&=(-)^n \int\Dint{w'} w'(t_1)\sdots w'(t_n) \PprobWSTAR[\wdpath] = (-)^n\WSTARave{w(t_1)\sdots w(t_n)}
\end{split}
\end{equation}
where we have introduced $w'(t)=-w(t)$ and used the $\SymParity$-symmetry in the equality leading to the second line. For odd $n$ \Eref{even_Pprob} implies $\WSTARave{w(t_1)\ldots w(t_n)}=0$. If all odd moments vanish, so do all odd cumulants, \ie all terms that $\corrStar_n$ and $W_n$ are comprised of vanish individually, rather than cancelling pairwise. As a result, $\nlowest$ must be even.

Finally, if $\Pprob^[\wpath]$ is $\SymTime$-symmetric, so that $\PprobW[\wpath]=\PprobW[\Rwpath]$ and $\EPRpartW=0$, \Eref{partial_EPW_defn_SM}, then all moments obey
\begin{equation}\elabel{reversible_Pprob}
\begin{split}
\WSTARave{w(T-t_1)\ldots w(T-t_n)}&=
\int\Dint{w} w(T-t_1)\sdots w(T-t_n) \PprobWSTAR[\wpath] = \int\Dint{w'} w'(t_1)\sdots w'(t_n) \PprobWSTAR[\Rwdpath]\\
&=\int\Dint{w'} w'(t_1)\sdots w'(t_n) \PprobWSTAR[\wdpath] = \WSTARave{w(t_1)\sdots w(t_n)}
\end{split}
\end{equation}
where we have introduced $w'(t)=w(T-t)$ and used the $\SymTime$-symmetry in the equality leading to the second line.  \Eref{reversible_Pprob} implies further that
$\WSTARaveS{w(t_1)\ldots w(t_n)}{\cbb}=\WSTARaveS{w(T-t_1)\ldots w(T-t_n)}{\cbb}$. As a result, for even $n$ in \Eref{def_corrStarETAL}, $\corrStar_n=0$ and $W_n=0$, so $\nlowest$ must be odd.

\section{Derivation of leading-order entropy production rates} \seclabel{leading_order}
In this section we present derivations of the leading-order EPRs of two $w$-processes discussed in the main text: the asymmetric telegraph process and diffusion with stochastic resetting.
The principal challenge is to determine the moments $\Wave{w(t_1)\sdots w(t_{\nlowest})}$ in each case as they are needed for the leading-order EPR, \Erefs{EPR_leading_order_main} and \eref{EPR_leading_order_SM}.

\subsection{Asymmetric Telegraph process}\seclabel{sm_asym_tel}
We consider the asymmetric telegraph process of $w(t)$, whereby the self-propulsion velocity switches between two fixed values $w(t) \in \{w_+, w_-\}$ with rate $\alpha_-$ from $w_+$ to $w_-$ and rate $\alpha_+$ from $w_-$ to $w_+$. As a Markov jump process between only two states, it is $\SymTime-$symmetric. As such, $\WSTARaveS{w(t_1)\ldots w(t_n)}{\cbb} = \WSTARaveS{w(T-t_1)\ldots w(T-t_n)}{\cbb}$, so the expansion \Eref{partial_EP_expansion_free} reduces to 
\begin{equation}\elabel{partial_EP_expansion_odd_terms}
    \EPRpartX 
    = \lim_{T \to \infty} \frac{1}{T} \sum_{n \textrm{ odd }}^\infty \frac{2}{n!}\left(\frac{\nu^2}{2D}\right)^n \int_0^T \dint{t_1}\!\!\sdots\dint{t_n} \Xave{\xdot(t_1)\sdots \xdot(t_n)}\,\WSTARaveS{w(t_1)\sdots w(t_n)}{\cbb}\;.
\end{equation}
If the process $w(t)$ is also $\SymParity$-symmetric, it will have zero odd moments and each term in the above expansion will therefore vanish. This is the case when  $w_+=-w_-$ and $\alpha_+=\alpha_-$. 
If the first moment $\Wave{w(t)}=(w_+\alpha_-+w_-\alpha_+)/(\alpha_++\alpha_-)$ does not vanish, the leading order is $\nlowest=1$, so that $\EPRpartX$ is given by, \Eref{asym_EPR_with_drift},
\begin{equation}\elabel{partial_EP_asym_telegraph_general}
    \EPRpartX 
    = \frac{\nu^2 \WaveS{w(t)}{2}}{D}+\OC\left(\frac{\nu^4}{D^2}\right)\;. 
\end{equation}
In the case of a drift-free particle, $\WSTARave{w(t)}=0$, the leading order partial EPR appears at third order in \Eref{partial_EP_expansion_odd_terms}, $\nlowest=3$, which we now calculate explicitly.

The process is described by the master equation
\begin{equation}
\frac{\plaind}{\plaind t}\begin{pmatrix}
p_+(t) \\
p_-(t) 
\end{pmatrix} =
\begin{pmatrix}
-\alpha_+ & \alpha_- \\
\alpha_+  & -\alpha_- 
\end{pmatrix}
\begin{pmatrix}
p_+(t) \\
p_-(t) 
\end{pmatrix}\;,
\end{equation}
where $p_+(t)$ is the probability of the process $w(t)$ taking the value $w_+$ at time $t$ and $p_-(t) = 1-p_+(t)$ is the probability that $w(t)$ takes the value $w_-$. Integrating yields the elements of a transition, \ie propagator matrix:
\begin{equation}\elabel{elements}
    \begin{split}
        P_{++}(t) &= \frac{1}{\alpha_+ + \alpha_-} \left(\alpha_- + \alpha_+ e^{-(\alpha_+ + \alpha_-)t}\right)\;, \\
        P_{-+}(t) &= \frac{1}{\alpha_+ + \alpha_-} \left(\alpha_+ - \alpha_+ e^{-(\alpha_+ + \alpha_-)t}\right)\;, \\
        P_{+-}(t) &= \frac{1}{\alpha_+ + \alpha_-} \left(\alpha_- - \alpha_- e^{-(\alpha_+ + \alpha_-)t}\right)\;, \\
        P_{--}(t) &= \frac{1}{\alpha_+ + \alpha_-} \left(\alpha_+ + \alpha_- e^{-(\alpha_+ + \alpha_-)t}\right)\;, \\
    \end{split}
\end{equation}
where $P_{ij}(t)$ is the probability of $w(t)$ taking the value $w_i$ having been initialised at $t=0$ with value $w_j$ where $i,j\in\{+,-\}$. Taking the limit $t\to\infty$ in \Erefs{elements} immediately yields the steady state probabilities. As we want to determine correlators of $w(t)$, we introduce the weighted steady state vector
\begin{equation}
    \wvec_0=\frac{1}{\alpha_++\alpha_-}
    \begin{pmatrix}
        w_+\alpha_-\\
        w_-\alpha_+
    \end{pmatrix}
\end{equation}
as well as $\bUnity=(1,1)^\transpose$, so that conveniently
\begin{equation}
    \Wave{w(t)}=
    \bUnity\cdot\wvec_0=
    \Wave{w} = 
    \frac{w_+\alpha_-+w_-\alpha_+}{\alpha_++\alpha_-}
\end{equation} 
independent of $t$ in the steady state. For now, we do not make use of $\Wave{w} =0$. In the spirit of a transfer matrix approach, we further introduce the weighted propagator matrix
\begin{gather}
    \Wmatrix(t) = 
    \begin{pmatrix}
        w_+ P_{++}(t) & w_+ P_{+-}(t)\\
        w_- P_{-+}(t) & w_- P_{--}(t)
    \end{pmatrix}
    =
    \Wmatrix_0 + \exp{-(\alpha_++\alpha_-)t}\Wmatrix_1\\
\text{where }\quad
\Wmatrix_0=
\frac{1}{\alpha_++\alpha_-}
\begin{pmatrix}
    w_+\alpha_- & w_+\alpha_-\\
    w_-\alpha_+ & w_-\alpha_+
\end{pmatrix}
=(\wvec_0 \wvec_0)
\quad\text{ and }\quad
\Wmatrix_1=
\frac{1}{\alpha_++\alpha_-}
\begin{pmatrix}
    w_+\alpha_+ & -w_+\alpha_-\\
    -w_-\alpha_+ & w_-\alpha_-
\end{pmatrix}\ .
\end{gather}
With a given time order, in the following $t_3\ge t_2 \ge t_1$,
correlation functions thus become a matter of some algebra, say
\begin{equation}
    \Wave{w(t_3)w(t_2)w(t_1)}=
    \bUnity \cdot
    \Wmatrix(t_3-t_2)\Wmatrix(t_2-t_1)
    \wvec_0
\end{equation}
which simplifies significantly by some helpful algebraic identities, in particular
\begin{subequations}
\begin{align}
    \bUnity^\transpose \Wmatrix_0 &= \Wave{w}\bUnity^\transpose\\
    \Wmatrix_0 \wvec_0 &= \Wave{w}\wvec_0\\
    \Wmatrix_0 \Wmatrix_0 &= \Wave{w} \Wmatrix_0 \ ,
\end{align}    
\end{subequations}
reflecting that $\Wmatrix_0=(\wvec_0 \wvec_0)$ is built from the stationary $\wvec_0$. Similarly
\begin{equation}
    \Wmatrix_1\Wmatrix_1 = 
    \frac{w_+\alpha_++w_-\alpha_-}{\alpha_++\alpha_-}\Wmatrix_1
    =\big(w_++w_--\Wave{w}\big)\Wmatrix_1
    \ ,
\end{equation}
so that 
\begin{multline}\elabel{3w_correlator_intermediate}
    \Wave{w(t_3)w(t_2)w(t_1)}=
\WaveS{w}{3}\\
+
\Big( 
\Wave{w} ( 
  \exp{-(\alpha_++\alpha_-)(t_3-t_2)} 
  + \exp{-(\alpha_++\alpha_-)(t_2-t_1)} - \exp{-(\alpha_++\alpha_-)(t_3-t_1)})  
+
(w_++w_-)\exp{-(\alpha_++\alpha_-)(t_3-t_1)}
\Big)
\bUnity \Wmatrix_1 \wvec_0
\ .
\end{multline}
For higher correlation functions, it is further useful that
\begin{equation}
    \Wmatrix_1\Wmatrix_0\Wmatrix_1
    = \frac{\alpha_+ \alpha_-(w_+ - w_-)^2}{(\alpha_+ + \alpha_-)^2}
    \Wmatrix_1 \ .
\end{equation}

By direct evaluation, we find
\begin{equation}
    \bUnity \cdot \Wmatrix_1 \wvec_0 = 
    \Wave{w^2}-\WaveS{w}{2}
\end{equation}
with steady state second moment 
$\Wave{w^2}=(w_+^2\alpha_-+w_-^2\alpha_-)/(\alpha_++\alpha_-)$, so that \Eref{3w_correlator_intermediate} is in principle fully evaluated. To proceed, we make use of the assumption $\Wave{w}=0$, so that now
\begin{equation}\elabel{3w_correlator_simplified}
    \Wave{w(t_3)w(t_2)w(t_1)}=
(w_++w_-)\exp{-(\alpha_++\alpha_-)(t_3-t_1)}\Wave{w^2}
\ .
\end{equation}
In principle, the leading order integral in \Eref{partial_EP_expansion_odd_terms} for $n=\nlowest=3$, \Erefs{EPR_leading_order_SM} and \eref{EPR_leading_order_main}, can now be carried out. To use \Eref{3w_correlator_simplified} without having to adjust the time-ordering, we use \Eref{simplified_integral} and write
\begin{multline}
\EPRpartX=
\lim_{T \to \infty} \frac{1}{T} 
\frac{2}{3!}\left(\frac{\nu^2}{2D}\right)^3 
(3!) \int_0^T \dint{t_1}\!\!\int_{t_1}^T\dint{t_2}\!\!\int_{t_2}^T\dint{t_3} 
\WaveS{w(t_3)w(t_2)w(t_1)}{2} + \OC\left(\frac{\nu^8}{D^4}\right)
\\
=
\quarter
\left(\frac{\nu^2}{D}\right)^3 (w_++w_-)^2 \WaveS{w^2}{2}
\frac{1}{4(\alpha_++\alpha_-)^2}
+ \OC\left(\frac{\nu^8}{D^4}\right)
\end{multline}
With $\Wave{w}=0$ the second moment simplifies to $\WaveS{w^2}{2}=w_+^2w_-^2$ and further $(w_++w_-)^2=-w_+w_-(\alpha_+-\alpha_-)^2/(\alpha_+\alpha_-)$, so that
\begin{equation}
\EPRpartX=
-\frac{1}{16} \left(\frac{\nu^2}{D}\right)^3 \frac{w_+^3w_-^3}{\alpha_+\alpha_-}\, \left(\frac{\alpha_+-\alpha_-}{\alpha_++\alpha_-}\right)^2
+ \OC\left(\frac{\nu^8}{D^4}\right) \;,
\end{equation}
which is the desired result \Eref{asym_EPR_without_drift}.

\subsection{Brownian motion with stochastic resetting}\seclabel{resetting_epr}

As the $w$-process we now consider a Brownian motion with diffusivity $\diffw$, subject to Poissonian resetting to $w=0$ with rate $\resettingRate$ \cite{Switkes:2004,PhysRevLett.106.160601}. Since the process has no bias it is $\SymParity-$symmetric, while $\SymTime-$symmetry is broken by the stochastic resetting. $\SymParity-$symmetry implies that odd moments and cumulants of $w(t)$ vanish so the expansion \Eref{partial_EP_expansion_free} reduces to 
\begin{equation}\label{eq:partial_EP_expansion_even_terms}
\begin{split}
    \EPRpartX 
    &= \lim_{T \to \infty} \frac{1}{T}\sum_{n \textrm{ even}}^\infty \frac{1}{n!}\left(\frac{\nu}{2D}\right)^n \int_0^T\dint{t_1}\!\!\sdots\dint{t_n} \Xave{\xdot(t_1)\sdots \xdot(t_n)}\left(\WSTARaveS{w(t_1)\sdots w(t_n)}{\cbb} - \WSTARaveS{w(T-t_1)\sdots w(T-t_n)}{\cbb}\right) \; . \\
    \end{split}
\end{equation}
The term at order $n=2$ vanishes due to the time-translation invariance of $w(t)$, \ie $\WSTARaveS{w(T-t_1) w(T-t_2)}{\cbb}=\WSTARaveS{w(t_1) w(t_2)}{\cbb}$, so the leading order partial EPR appears at order $\nlowest=4$, \Erefs{EPR_leading_order_SM} and \eref{EPR_leading_order_main}. In the following, we explain how to calculate higher-order correlation functions of this process and provide an explicit expression of the four-time correlation function.

Correlation functions are readily expressed in terms of convolutions of the propagator $\propagator{w,t;w_0,t_0}$, which is the probability density of finding the particle at $w$ at time $t$, after placing it at $w_0$ at time $t_0$. Not only are these convolutions more easily carried out in $k$-space, this representation consists in the characteristic function \cite{vanKampen:1992} and therefore lends itself naturally to calculating the correlation functions. We therefore introduce 
\begin{equation}
\propagator{w,t;w_0,t_0}
=
\int\dintbar{k_0}\dintbar{k}
\exp{\imag k_0 w_0} \exp{\imag k w}
\propagatorFourier{k,t;k_0,t_0}
\end{equation}
where $\dbar k=\plaind k/(2\pi)$
and
$\propagatorFourier{k,t;k_0,t_0}$ is the propagator in $k,t$, straight-forwardly characterised \cite{Maziya:2023} using Doi-Peliti field theory \cite{Doi:1976,Peliti:1985}
\begin{equation}\elabel{propagator_field_theory}
    \propagatorFourier{k_1,t_1;k_0,t_0}=\deltabar(k_1+k_0) \theta(t_1-t_0) \exp{-(\diffw k_1^2 + \resettingRate)(t_1-t_0)}
    + \theta(t_1-t_0)\frac{\resettingRate\deltabar(k_0)}{\diffw k_1^2 + \resettingRate} \left(
    1-\exp{-(\diffw k_1^2 + \resettingRate)(t_1-t_0)}\right)
\end{equation}
where $\deltabar(k+k_0)=2\pi\delta(k+k_0)$ and $\theta(t_1-t_0)$ is the Heaviside $\theta$-function. In \Eref{propagator_field_theory} times $t_n$ are indexed suggestively, and we will indeed assume $t_n\ge t_{n-1}$ in the following. To ease notation, we introduce 
\begin{subequations}
\begin{align}
    a_n(k)&=\exp{-(\diffw k^2 + \resettingRate)(t_n-t_{n-1})}\\
    b_n(k)&=\frac{\resettingRate}{\diffw k^2 + \resettingRate}
    \left(1-\exp{-(\diffw k^2 + \resettingRate)(t_n-t_{n-1})}\right)
\end{align}
\end{subequations}
so that 
\begin{equation}\elabel{propagator_field_theory_in_ab}
    \propagatorFourier{k_1,t_1;k_0,t_0}=
    \deltabar(k_1+k_0) \theta(t_1-t_0) a_1(k_1)
    +
    \deltabar(k_0) \theta(t_1-t_0) b_1(k_1) \ .
\end{equation}
To illustrate the scheme to be used in the following, we calculate the one-point function 
\begin{equation}
    g(t_1,t_0)=
    \int\dint{w_1}w_1\int\dintbar{k_{1,0}}
    \exp{\imag k_1 w_1}\exp{\imag k_0 w_0}\propagatorFourier{k_1,t_1;k_0,t_0} \ ,
\end{equation}
where the integral $\int\dintbar{k_{1,0}}\sdots$ is over all $k_1,k_0\in\Rset$. It is easily evaluated by expressing $w_1$ as $-\imag\plaind/\plaind k_1$, integrating by parts about $k_1$, taking the integral over $w_1$ and using the resulting $\deltabar(k_1)$ to carry out the integral over $k_1$,
\begin{equation}
    g(t_1,t_0)=\imag\int\dintbar{k_0} \exp{\imag k_0 w_0} 
    \left.\frac{\plaind}{\plaind k_1}\right|_{k_1=0} 
    \propagatorFourier{k_1,t_1;k_0,t_0}\ .
\end{equation}
This integral is easy to perform, after another integration by parts finally producing 
\begin{equation}
    g(t_1,t_0)=
    w_0\exp{-\resettingRate (t_1-t_0)} \ .
\end{equation}

Throughout the present work, path probabilities are taken in the steady state. In the present framework, this has to be implemented explicitly by taking $t_0\to-\infty$. Using the notation of \Eref{def_bave}, $\lim_{t_0\to-\infty}g(t_1,t_0)=\Wave{w(t_1)}=0$ and similarly
\begin{subequations}
\begin{align}
    \Wave{w(t_2)w(t_1)}
    &=  \lim_{t_0\to-\infty} \int\dint{w_{1,2}} w_1 w_2 \int\dintbar{k_{2,1',1,0}}
    \exp{\imag k_2w_2}\exp{\imag k_{1'}w_1}\exp{\imag k_1w_1}\exp{\imag k_0w_0}
    \propagatorFourier{k_2,t_2;k_{1'},t_1}
    \propagatorFourier{k_1,t_1;k_0,t_0}\\
    \elabel{ww_secondline}
    &= \imag^2 \int \dintbar{k_{2,1',1,0}} \deltabar(k_2) \deltabar(k_{1'}+k_1) \exp{\imag k_0w_0} 
    \frac{\plaind}{\plaind k_2} \left(\deltabar(k_2+k_{1'}) a_2(k_2)+\deltabar(k_{1'}) b_2(k_2)\right) \\
    &\quad\times\lim_{t_0\to-\infty} \frac{\plaind}{\plaind k_1} \Big(\deltabar(k_1+k_{0}) a_1(k_1)+\deltabar(k_{0}) b_1(k_1)\Big)\nonumber 
\end{align}
\end{subequations}
where we have introduced the useful notation of dashed indices indicating Fourier-modes $k_{n'}$ when particles carry on propagating after having been observed. The limit is easily carried out, 
\begin{equation}
    \lim_{t_0\to-\infty}
    \propagatorFourier{k_1,t_1;k_0,t_0} =
    \lim_{t_0\to-\infty}
    \Big(\deltabar(k_1+k_{0}) a_1(k_1)+\deltabar(k_{0}) b_1(k_1)\Big)
    =\deltabar(k_0)  \frac{\resettingRate}{\diffw k_1^2 + \resettingRate} =: \deltabar(k_0) \bbar(k_1)
\end{equation}
reproducing the expected Lorentzian $\bbar(k_1)$ in the steady state \cite{PhysRevLett.106.160601}. 
Proceeding from \Eref{ww_secondline} in this way,
\begin{subequations}
\begin{align}
    \Wave{w(t_2)w(t_1)}
    &=  - \int \dintbar{k_{1',1}} \deltabar(k_{1'}+k_1)
    \left.\frac{\plaind}{\plaind k_2}\right|_{k_2=0} \Big(\deltabar(k_2+k_{1'}) a_2(k_2)+\deltabar(k_{1'}) b_2(k_2)\Big) \bbar'(k_1)\\
    &=\ldots=-a_2(0) \bbar''(0)
\end{align}
\end{subequations}
where derivatives are indicated by dashes. Derivatives of $\delta$-functions are interpreted by a suitable integration by parts. As $a_2(0)=\exp{-\resettingRate(t_2-t_1)}$ and $\bbar''(0)=-2\diffw/\resettingRate$, the desired correlation function follows,
\begin{equation}
    \Wave{w(t_2)w(t_1)} = \frac{2\diffw}{\resettingRate}\exp{-\resettingRate|t_2-t_1|} \ ,
\end{equation}
where we have lifted the constraint $t_2>t_1$ by introducing a modulus.

Similarly, the four-point function
\begin{align}
    \Wave{w(t_4)w(t_3)w(t_2)w(t_1)} &=
    (\imag)^4 \int\dintbar{k_{4,3',3,2',2,1',1,0}}
    \deltabar(k_4)\deltabar(k_{3'}+k_3)\deltabar(k_{2'}+k_2)\deltabar(k_{1'}+k_1)\exp{\imag k_0w_0}\\
    &\times\frac{\plaind}{\plaind k_4} \Big(\deltabar(k_4+k_{3'})a_4(k_4)+\deltabar(k_{3'})b_4(k_4)\Big)\nonumber\\
    &\times\frac{\plaind}{\plaind k_3} \Big(\deltabar(k_3+k_{2'})a_3(k_3)+\deltabar(k_{2'})b_3(k_3)\Big)\nonumber\\
    &\times\frac{\plaind}{\plaind k_2} \Big(\deltabar(k_2+k_{1'})a_2(k_2)+\deltabar(k_{1'})b_2(k_2)\Big)\nonumber\\
    &\times\frac{\plaind}{\plaind k_1} \Big(\phantom{\deltabar(k_2+k_{1'})a_2(k_2)\;\;+}\deltabar(k_{0})\bbar(k_1)\Big)\nonumber
\end{align}
which produces
\begin{equation}\elabel{w4}
    \Wave{w(t_4)w(t_3)w(t_2)w(t_1)} = 
    \exp{-\resettingRate (t_4-t_1)}
    \left(\frac{20\diffw^2}{\resettingRate^2} + 12 \frac{\diffw^2}{\resettingRate}(t_2-t_1)\right)
    + \exp{-\resettingRate(t_4-t_3+t_2-t_1)} \frac{4\diffw^2}{\resettingRate^2}
\end{equation}
under the constraint $t_4>t_3>t_2>t_1$. \Eref{w4} 
immediately produces the connected correlation function via
\begin{multline}
    \WaveS{w(t_4)w(t_3)w(t_2)w(t_1)}{\cbb}
    =
    \Wave{w(t_4)w(t_3)w(t_2)w(t_1)} \\
    - \Wave{w(t_4)w(t_3)}\,\Wave{w(t_2)w(t_1)}
    - \Wave{w(t_4)w(t_2)}\,\Wave{w(t_3)w(t_1)}
    - \Wave{w(t_4)w(t_1)}\,\Wave{w(t_3)w(t_2)}
\end{multline}
namely
\begin{equation}\elabel{w4c}
\WaveS{w(t_4)w(t_3)w(t_2)w(t_1)}{\cbb}
    =
    \exp{-\resettingRate (t_4-t_1)}
    \left(\frac{20\diffw^2}{\resettingRate} + 12 \frac{\diffw^2}{\resettingRate}(t_2-t_1)\right)
    + \exp{-\resettingRate(t_4+t_3-t_2-t_1)} \frac{8\diffw^2}{\resettingRate^2}\ ,
\end{equation}
which differs from \Eref{w4} only by the pre-factor and the argument in the exponential in the last term. 

In order to calculate the EPR, it is necessary to evaluate \Erefs{w4} and \eref{w4c} at times $T-t_i$, which gives
\begin{multline}\elabel{corr_difference}
    \Wave{w(t_4)w(t_3)w(t_2)w(t_1)}
    -
    \Wave{w(T-t_1)w(T-t_2)w(T-t_3)w(T-t_4)} 
= 
12 \frac{\diffw^2}{\resettingRate}
\big((t_2-t_1)-(t_4-t_3)\big)
\exp{-\resettingRate (t_4-t_1)}\\
=
    \WaveS{w(t_4)w(t_3)w(t_2)w(t_1)}{\cbb}
    -
    \WaveS{w(T-t_1)w(T-t_2)w(T-t_3)w(T-t_4)}{\cbb}
\end{multline}
written in the suggestive form as $T-t_1>T-t_2>T-t_3>T-t_4$
iff $t_4>t_3>t_2>t_1$, even when the factors of $w(t_i)$, of course, commute. 

\subsubsection{Entropy production rate}\seclabel{EPR_resetting}
In the form \Eref{simplified_integral}, the integral \Eref{EPR_leading_order_SM} can be carried out by any computer algebra system, producing
\begin{align}\elabel{EPR_resetting_key_integral_rewritten}
    \EPRpartX 
    &= \lim_{T \to \infty} \frac{1}{2T} 
    \left(\frac{\nu^2}{2D}\right)^4 
        \int_0^T \dint{t_1}
    \int_{t_1}^T \dint{t_2}
    \int_{t_2}^T \dint{t_3}
    \int_{t_3}^T \dint{t_4}
\left(\frac{12 \diffw^2}{\resettingRate}\right)^2
\big((t_2-t_1)-(t_4-t_3)\big)^2
\exp{-2\resettingRate (t_4-t_1)}
 + \OC\left( \nu^{10}D^{-5} \right)\nonumber \\
 &= \frac{9}{32}\left(\frac{\nu^2}{D}\right)^4\frac{\diffw^4}{\resettingRate^7} 
 + \OC\left( \nu^{10}D^{-5} \right) \;,
\end{align}
which is the desired result, \Eref{resetting_epr}.

\end{document}